\documentclass{article}
\usepackage{arxiv}

\usepackage{microtype}                 % use micro-typography (slightly more compact, better to read)
\PassOptionsToPackage{warn}{textcomp}  % to address font issues with \textrightarrow
\usepackage{textcomp}                  % use better special symbols
\usepackage{mathptmx}                  % use matching math font
\usepackage{times}                     % we use Times as the main font
\usepackage{cite}                      % needed to automatically sort the references
\usepackage{tabu}                      % only used for the table example
\usepackage{booktabs}                  % only used for the table example

\usepackage{subfig} % for subfigure
\usepackage{amsmath} % to delete equation number
\usepackage{amssymb} % complexity
\usepackage{algorithm} % pseudocode
\usepackage{algorithmic} % pseudocode
\usepackage{graphicx}
\usepackage{url}            % simple URL typesetting

\title{Orthogonal Voronoi Diagram and Treemap}

\author{
Yan-Chao Wang \\
School of Computer Science and Engineering \\
Nanyang Technological University \\
Singapore \\
\texttt{ywang054@e.ntu.edu.sg} \\
\And
Feng Lin\\
School of Computer Science and Engineering \\
Nanyang Technological University \\
Singapore \\
\texttt{asflin@ntu.edu.sg} \\
\And
Hock-Soon Seah\\
School of Computer Science and Engineering \\
Nanyang Technological University \\
Singapore \\
\texttt{ashsseah@ntu.edu.sg} \\
}

\begin{document}
\maketitle

%% Abstract section.
\begin{abstract}
In this paper, we propose a novel space partitioning strategy for implicit hierarchy visualization such that the new plot not only has a tidy layout similar to the treemap, but also is flexible to data changes similar to the Voronoi treemap. To achieve this, we define a new distance function and neighborhood relationship between sites so that space will be divided by axis-aligned segments. Then a sweepline+skyline based heuristic algorithm is proposed to allocate the partitioned spaces to form an orthogonal Voronoi diagram with orthogonal rectangles. To the best of our knowledge, it is the first time to use a sweepline-based strategy for the Voronoi treemap. Moreover, we design a novel strategy to initialize the diagram status and modify the status update procedure so that the generation of our plot is more effective and efficient. We show that the proposed algorithm has an $O(n \cdot \text{log}(n))$ complexity which is the same as the state-of-the-art Voronoi treemap. To this end, we show via experiments on the artificial dataset and real-world dataset the performance of our algorithm in terms of computation time, converge rate, and aspect ratio. Finally, we discuss the pros and cons of our method and make a conclusion.

\end{abstract} % end of abstract

%% Keywords that describe your work. Will show as 'Index Terms' in journal
%% please capitalize first letter and insert punctuation after last keyword
\keywords{Orthogonal rectangle \and Voronoi diagram \and Voronoi treemap\and hierarchy visualization\and treemap\and axis-aligned rectangle}

\section{Introduction}

%% \section{Introduction} %for journal use above \firstsection{..} instead
Hierarchical data visualization plays an essential role in the field of information visualization since hierarchical data structures are quite common in our daily life, including file systems, software/package class structures, and the organization structure of companies. Thus, a large number of hierarchy visualization methods have been developed to depict the dataset from different aspects~\cite{Graham:10, Schulz:11}. Some of them explicitly show the hierarchical structure as straight lines, arcs, or curves, while others focus on the value within each node and positionally encode the hierarchy by node overlap or inclusion. Hence, these methods implicitly presenting the hierarchy are considered more space efficient~\cite{Schulz:11a}. Treemap~\cite{Shneiderman:92} and Voronoi treemap~\cite{Balzer:05} which are formed by nested rectangles and polygons respectively are the most two popular implicit hierarchy visualization methods. However, they adopt distinct strategies to generate the plots. Treemap method divides the empty canvas into rectangular sub-regions so that the area is associated with the relative sizes of the respective sub-hierarchies. Meanwhile, Voronoi treemap partitions the canvas into polygon shapes based on distance to beforehand specified sites. After that, the position and weight of the sites may be iteratively adapted in order to adjust the area of the sub-regions.

In our opinion, the treemap and the Voronoi treemap make different choices in terms of shape simplicity and adjustment ability. To be specific, by using rectangular shapes, the treemap is more comfortable and much tidier for the viewers than Voronoi treemap with nested polygons. Moreover, rectangle shapes make it easier to visually compare the areas of different regions than polygon shapes ~\cite{Duarte:14}. However, when the hierarchical data changes even slightly, the partitioning of empty canvas in treemap need to be regenerated and the new layout may be largely different from the previous one. It usually leads to bad layout stability. In contrast, Voronoi treemap can adjust its layout via slight modification on the status of its sites. 

We aim to find a good balance between shape simplicity and adjustment ability with the proposed novel hierarchy visualization method: orthogonal Voronoi diagram and treemap. The orthogonal layout has been proved to suitable for human beings in explicit hierarchy visualization (node-link diagram) and network visualization~\cite{Burch:11, Kieffer:16}. It would be an interesting topic to evaluate different treemap layouts in implicit hierarchy visualization. A treemap with orthoconvex and L-shape was proposed by dividing existing treemap element to preserve an aspect ratio constraint~\cite{De:14}. However, the orthoconvex treemap cannot be adjusted flexibly. 

\begin{figure}
  \centering
  \subfloat[Orthogonal Voronoi Diagram]{{\includegraphics[width= 0.31 \linewidth]{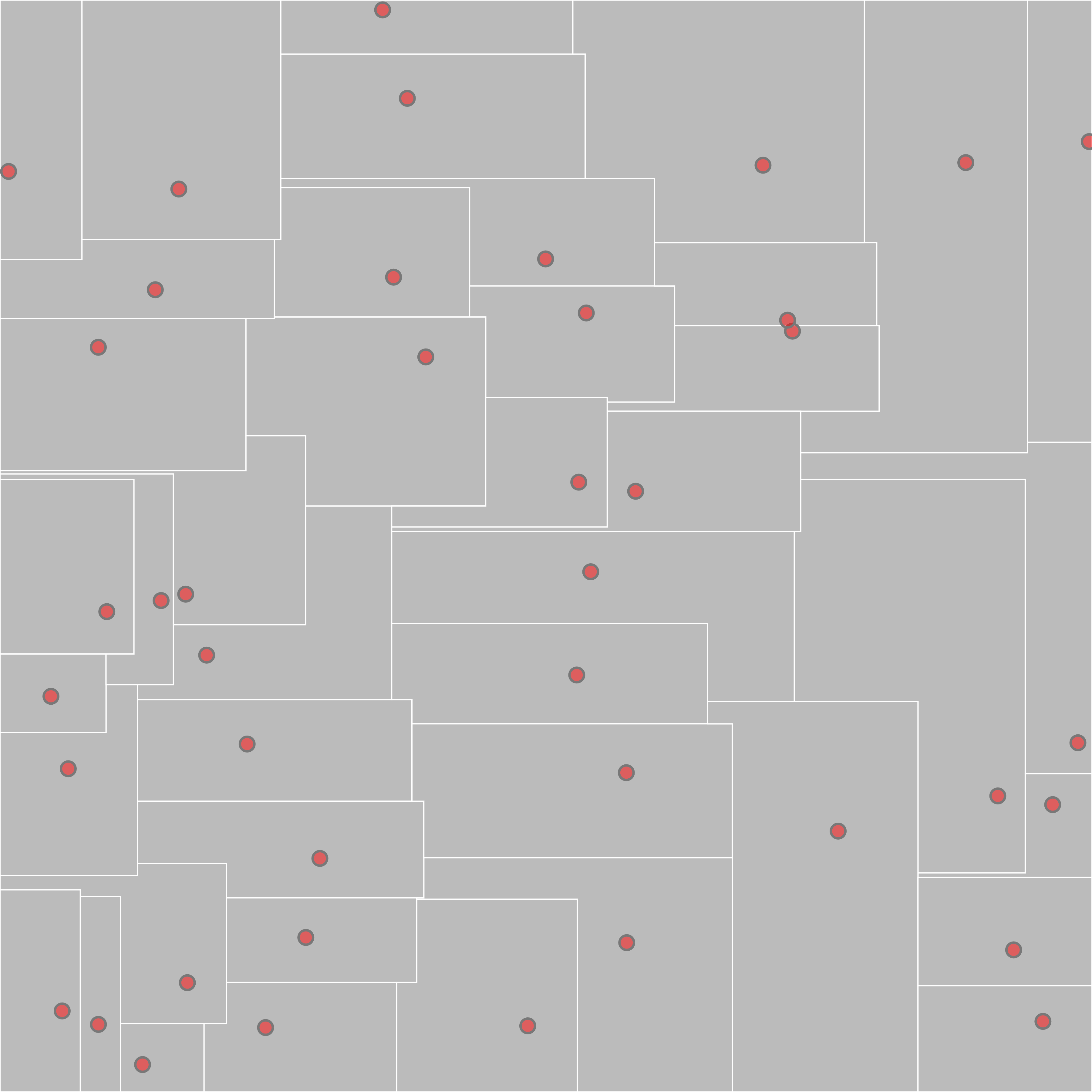} }} 
  \quad
  \subfloat[Orthogonal Voronoi Treemap (Random)]{{\includegraphics[width= 0.31 \linewidth]{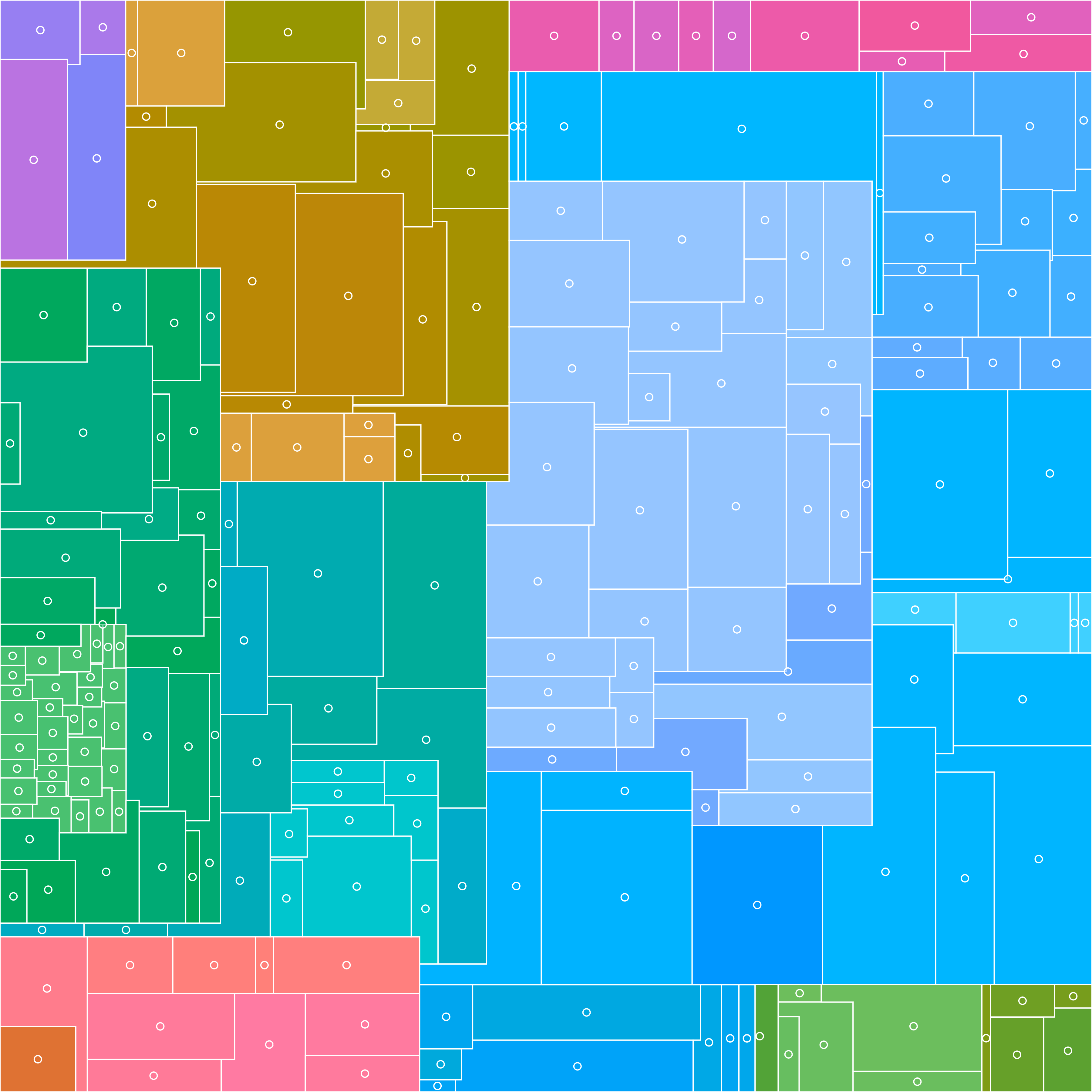} }} 
  \quad
  \subfloat[Orthogonal Voronoi Treemap ]{{\includegraphics[width= 0.31 \linewidth]{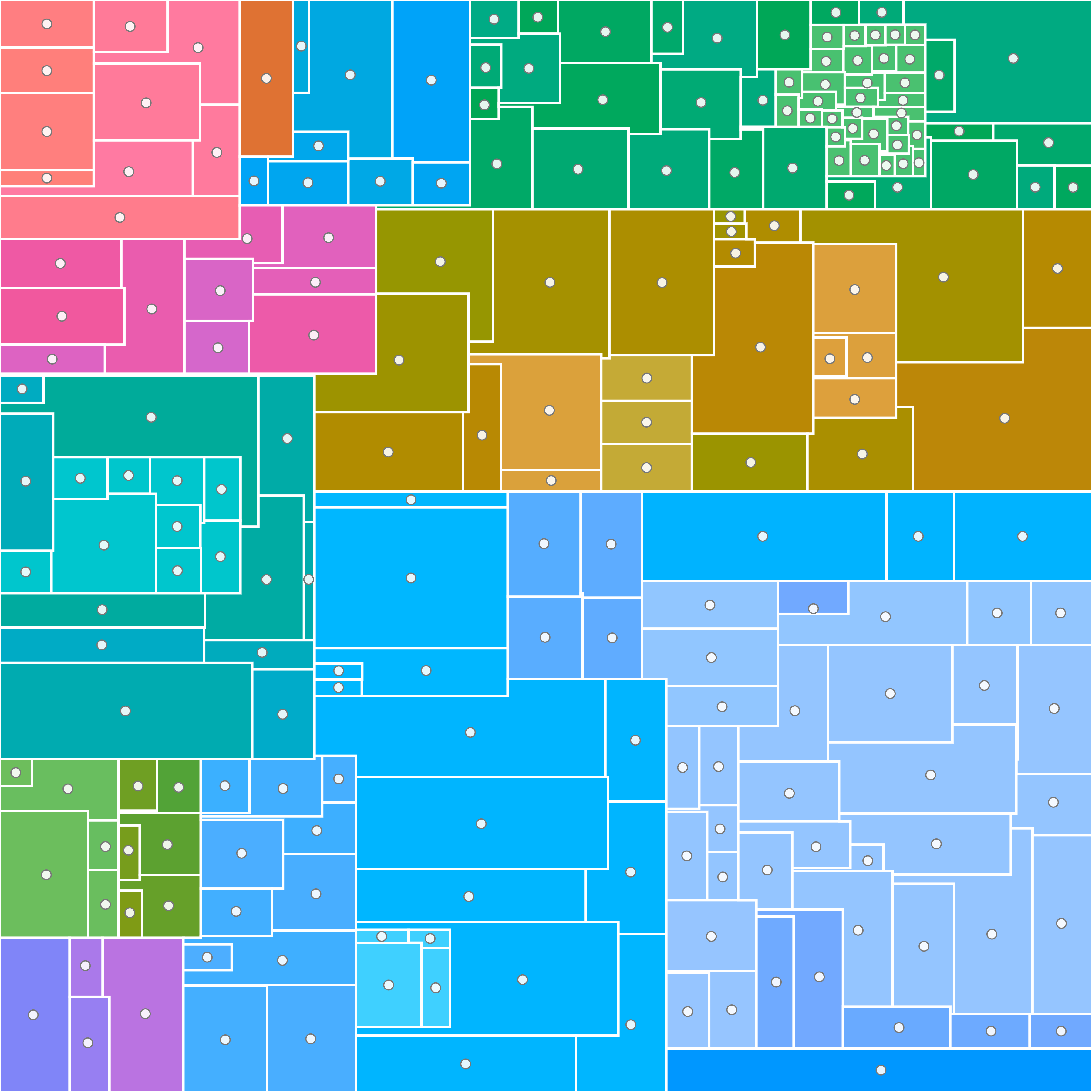} }}
  \caption{Orthogonal Voronoi diagram and orthogonal Voronoi treemap. (a) The orthogonal Voronoi diagram is generated based on the positions of a series of random points. (b) The orthogonal Voronoi treemap visualizes the Flare class hierarchy (colored by TreeColors~\cite{Tennekes:14}) with random initial status. (c) The orthogonal Voronoi treemap uses the squarified treemap as initial status.  
  }
  \label{fig:teaser}
\end{figure}

Our orthogonal Voronoi diagram partitions the empty canvas into nested orthogonal rectangles. Hence, the generated layout is not only flexible to a diversified data value, but also much tidier than the Voronoi treemap with nested polygons. To achieve this, we first define a new distance calculation strategy in order to generate axis-aligned segmentation among the sites. In the strategy, a new distance function is defined by considering the relative positions of two sites rather than a single site. Then a horizontal or vertical segmentation line is generated between these two sites. After that, we design a sweepline + skyline heuristic algorithm to partition the canvas based on the new distance calculation strategy to generate an orthogonal Voronoi diagram. The proposed algorithm is motivated by the sweep line algorithm for Voronoi diagram generation~\cite{Fortune:87} and the skyline strategy in handling cutting and packing problems~\cite{Burke:04}. By iteratively alerting the status of the original sites and calling the sweepline + skyline algorithm, the area of the orthogonal rectangular sub-regions will match their corresponding values. An orthogonal Voronoi treemap will be obtained if this process is recursively continued layer by layer until the whole hierarchical structure is traversed. Moreover, we also design an initialization strategy to improve algorithm performance. The proposed algorithm is fast, simple, and resolution-independent. Figure~\ref{fig:teaser} gives examples of the proposed orthogonal Voronoi diagram (on a series of random sites marked in red with zero weight) and the orthogonal Voronoi treemap (of the hierarchical structure of the Flare classes, starting from different initial statuses).

%% Uncomment below to include a teaser figure.

The main contribution of this paper is threefold: First, we define a novel distance function based on two sites so that the segmentation is axis-aligned. Second, we design a sweepline + skyline algorithm for space partitioning with $O(n \cdot \text{log} (n))$ complexity. To the best of our knowledge, this is the first time the sweep line strategy is used for the Voronoi treemap. Third, we design a novel strategy to initialize the diagram status and modify the status update procedure to increase the efficiency and effectiveness of our algorithm.

The rest of this paper is organized as follows. The related work about our orthogonal Voronoi treemap is reviewed in Sect.~\ref{sec:relatedwork}. The background knowledge is introduced in Sect.~\ref{sec:background} before the description of our methodology. In Sect.~\ref{sec:treemap}, we describe the proposed orthogonal space partitioning algorithm to generate the orthogonal Voronoi treemap. The performance of our algorithm is depicted in Sect.~\ref{sec:performance}. Finally, discussion and conclusion are made in Sect.~\ref{sec:discuss} and Sect.~\ref{sec:conclusion}.

\section{Related Work}
\label{sec:relatedwork}
In this section, we give an overview of implicit hierarchy visualization methods which are related to our work. We mainly focus on the canvas subdivision strategies used to generate layouts, instead of including all implicit visualization techniques. Based on whether the sites are referred to during the subdivision, we divide the methods into two clusters: non-site-based methods and site-based methods. Both of them belong to methods with inclusion edge representation according to the design space definition~\cite{Schulz:11a}. 

\paragraph{Non-site-based Methods} Implicit hierarchy visualization methods that partition the whole space without considering the sites are treated as non-site-based methods, such as the treemap. These methods position the data by following some rules or experience in order to get expected configurations, which sometimes are also named as heuristic-based algorithm. Starting from the propose of original treemap in 1992~\cite{Shneiderman:92}, a large number of variants are proposed in the literature~\cite{Graham:10, Wang:16}.

The squarified treemap focuses on the emergence of thin, elongated rectangles in the standard treemaps and presents a new subdivision method such that the resulting rectangles have a lower aspect ratio~\cite{Bruls:00}. The ordered treemap layout is the first type of treemap layout that takes stability into consideration~\cite{Shneiderman:01}. In their work, two pivot based algorithms (pivot-by-size and pivot-by-middle) are proposed to ensure that items near each other in the original data will be near each other in the final layout. The split algorithm used in the ordered and quantum treemaps~\cite{Bederson:02} is a modification of the squarified treemaps, following a given one-dimensional ordering. The spiral treemap positions the one-dimensional ordering of the input data along the border following a circular arrangement or an S-shape~\cite{Tu:07}. Different from previous methods which only consider one dimension, the spatially order treemaps consider two-dimensional consistency by relating node order to Euclidean distance from the parent node's top-left corner~\cite{Wood:08}. We observe that the layout generation problem in treemap is quite similar to the two-dimensional (2D) bin packing which is an optimization problem with a wide range of applications in resource management. This observation was also mentioned by Schulz et al.~\cite{Schulz:11a}. Since many heuristic algorithms~\cite{Burke:04, Wang:15} have been proposed to solve the bin packing problem, how to utilize them into the layout generation in treemap would be an interesting research direction and we find that some researchers have started to do this~\cite{Itoh:04, Kobayashi:12}.

Non-rectangular treemaps are also designed in the literature. Jigsaw map has nicely shaped regions and stable layout by considering Hilbert curves or H curves~\cite{Wattenberg:05}. They generate irregular shapes which are not easy to be compared with. A modification then was proposed by splitting the space into rectangles~\cite{Tak:13}. To relax rectangular constraint, angular treemaps describe a divide-and-conquer method to partition the space into various shapes~\cite{Liang:12}. Besides that, the treemap layout that produces irregular nested shapes by subdividing the Gosper curve~\cite{Auber:13} was also proposed.

\paragraph{Site-based Methods} Some implicit hierarchy visualization methods partition the space based on a series of pre-defined sites, such as the Voronoi treemap. The Voronoi treemap was originally presented by Balzer et al.~\cite{Balzer:05}. By relaxing the constraint of rectangular shapes, they utilize Voronoi tessellations to generate polygonal subdivisions. They firstly initialize a set of sites with initial weight values and then compute the Voronoi tessellations based on distance functions. By adaptively altering the weight value of each site, it enables a dedicated Voronoi region in the next iteration step. Finally, the computation will be stopped when a good enough layout is reached. Later, the Voronoi treemaps are utilized to visualize dynamic hierarchical data owing to its adjustment ability~\cite{Sud:10, Gotz:11}. However, the calculation of these Voronoi treemaps is computationally expensive as a random-sampling strategy is used to compute the Voronoi tessellations. In 2012, Nocaj and Brandes~\cite{Nocaj:12} proposed a resolution-independent algorithm by calculating the Voronoi tessellations with power diagrams, such that the new algorithm is faster in both theory and practice. An improvement is then made by setting an initial position for visualizing varying hierarchies~\cite{Hahn:14}.

Neighborhood treemap (Nmap)~\cite{Duarte:14}, that successively bisects a set of pre-defined sites on the horizontal or vertical directions and then scales the bisections to match the value of each site, is also a site-based method. Although no distance function is used during the segmentation, Nmap also needs sites representing the similarity relationships of data elements to be positioned in the canvas. Thus, Nmap can preserve similarity relationship among data elements very well. However, no evidence shows that Nmap can produce stable layouts with dynamic data. Circle packing~\cite{Wang:06} can also be treated as a site-based method since the generation of the layouts is based on the center of each circle, as well as the recently proposed bubble treemaps~\cite{Gortler:18}.

\section{Background}
\label{sec:background}
In this section, we review the background on the Voronoi treemap, including Voronoi diagram, weighted Voronoi diagram, centroidal Voronoi diagram, and Voronoi treemap. In the description, we follow the notation used in Nocaj and Brandes's work~\cite{Nocaj:12}.

\subsection{Voronoi Diagram}
A Voronoi diagram (also called a Voronoi tessellation, or a Voronoi partition) is a partitioning of a plane into sub-regions based on distances to a set of points within the plane. These sub-regions are often called \textit{Voronoi cell} (or \textit{cell}) and these points in the plane are called \textit{sites}. In this paper, we only consider the partitioning in a bounded region (e.g. a rectangle) rather than the whole 2D plane.

Formally, given a bounded region $\Omega \subset R^{2}$ and a set of $n$ sites $S= \{s_1, s_2, ..., s_n\}$, the Voronoi diagram divides $\Omega$ into a set of Voronoi cells $\upsilon(s_i)$, one for each site $s_i$. Then the cell $\upsilon(s_i)$ can be expressed as
\begin{equation}
\label{eq:cell}
\upsilon(s_i) = \{p\in\Omega \ | \ dist( p,s_{i}) < dist( p,s ). \ \forall s\in S, s\neq s_{i} \},
\end{equation}
where $dist(p,s_i)$ is the distance between point $p=(x_p, y_p)$ and site $s_i=(x_{s_i}, y_{s_i})$. The distance can be the Euclidean distance or other distance functions. The most two often used distance functions are the Euclidean distance $dist_{e}(p, s_i)$ and the power Euclidean distance $dist_{pe}(p, s_i)$ which are shown as below:
\begin{equation}
dist_{e}(p, s_i) = \|p-s_i\| = \sqrt {\left( x_{p}-x_{s_i}\right) ^{2}+\left( y_{p}-y_{s_i}\right) ^{2}},
\end{equation}
\begin{equation}
\label{eq:pe}
dist_{pe}(p, s_i) = \|p-s_i\|^{2} = \left( x_{p}-x_{s_i}\right) ^{2}+\left( y_{p}-y_{s_i}\right) ^{2}.
\end{equation}

Thus, the Voronoi diagram is defined as the collection of Voronoi cells, $\upsilon(S) = \{ \upsilon(s_1), ... , \upsilon(s_n) \}$.

\subsection{Weighted Voronoi Diagram}
In the Voronoi diagram, the area of a cell is fixed and only depends on the positions of its associated and neighboring sites. Hence, in order to use the cells to depict additional information (e.g. data value), a mechanism to control the areas of cells is required. To achieve this, a positive real weight is associated with each site. When calculating the distance, the associated weights should also be taken into account. 

Since the power Euclidean distance function produces a straight line between two sites (the solid black lines in Fig.~\ref{fig:dist1}), thus in this paper the Voronoi plots are all based on the power Euclidean distance with the exception of our proposed orthogonal Voronoi plots that is based on a new distance function. 

Formally, let $W = \{w_1, w_2, ... , w_n\} $ be a set of positive weights associated with the set of sites $S$ correspondingly. Then the weighted power Euclidean distance based on the original power Euclidean distance (Eq.~\ref{eq:pe}) can be written as follows:
\begin{equation}
dist_{awpe}(p, s_i) = \|p-s_i\|^{2} - w_i.
\end{equation}

Increasing the weight value will increase the area of the cell. However, it is nonlinear in general. Besides that, a too large weight value may lead to empty cells. We will discuss this overweight issue in the description of our algorithms.

\subsection{Centroidal Voronoi Diagram}
Centroidal Voronoi diagram is a special type of Voronoi diagram that the site $s_i$ is located at the center of each cell $v(s_i)$~\cite{Du:99}. This kind of diagram usually can generate cells with good aspect ratio (i.e. the ratio of the sides of the oriented minimum bounding rectangle is close to one). This is relevant because a good aspect ratio ensures a good readability in visualization. Let $\overline{\upsilon(s_i)}$ and $A(\upsilon(s_i))$ be the polygonal boundary and area of cell $\upsilon(s_i)$ respectively, the centroid $c_i = centroid(\overline{\upsilon(s_i)})$ can be calculated in linear time. For all the $n$ sites, the computation of the centroid needs $O(n)$ time.

\subsection{Voronoi Treemap}
The Voronoi treemap is a recursive partitioning of a plane, the same as the treemap. Starting from the root of a hierarchy, a weighted Voronoi diagram is generated in the region $\Omega$ with one cell for each child of the root.  An iterative optimization process is taken to adaptively alert the value of weight and the position of the site, such that the areas of the cells meet the requirement. The final layout requires that the area of each cell should be in proportion to the associated value. In practice, if the area error is smaller than a threshold, then we will say the requirement is met and the iterative process is converged. Formally, let $value_{s_i}$ be the associated value of site $s_i$ and $E_{threshold}$ be the threshold of the area error, then the convergent requirement can be expressed as:
\begin{equation}
\label{eq:converge}
\dfrac {\sum _{s_{i}\in S}\left| A\left( \upsilon\left( s_{i}\right) \right) -A\left( \Omega \right) * \dfrac {value_{s_{i}}}{value_{S} }\right| }{A\left( \Omega \right) } < E_{threshold}.
\end{equation}
Once the iterative process is converged, the above-mentioned processes recurse to subdivide child region until all the leaves of the hierarchy are represented by cells with desired areas.

\section{Orthogonal Voronoi Treemap}
\label{sec:treemap}
In this section, we introduce our orthogonal Voronoi treemap algorithm (OVT). We first provide an overview of our algorithm in Sect.~\ref{subsec:overview}, and then we discuss the three main steps of our algorithm: initialize site status (Sect.~\ref{subsec:initialization}), update site status (Sect.~\ref{subsec:update}), and compute the weighted orthogonal Voronoi diagram (Sect.~\ref{subsec:diagram}). Finally, we give some implementation details in Sect.~\ref{subsec:implementation}.

The overall structure of our algorithm follows that of Nocaj and Branches~\cite{Nocaj:12}. Meanwhile, ours is different in the following aspects:
%$\dot{•}$ a new initialization strategy;\\
%$\dot{•}$ a new initialization strategy;\\
%$\dot{•}$ a new initialization strategy;\\
%$\dot{•}$ a new initialization strategy;\\

\begin{itemize}
	\item a new initialization strategy;
	\item a modified site update strategy;
	\item a novel distance function;
	\item a heuristic algorithm for diagram generation;
\end{itemize} 

\subsection{Overview}
\label{subsec:overview}

In this paper, the proposed orthogonal Voronoi treemap follows the same rules as the traditional Voronoi treemap. Because of the recursive feature of treemaps, considering a single layer is sufficiently enough. A pseudo-code to compute single layer orthogonal Voronoi treemap is summarized in Algorithm~\ref{alg:treemap}.

\begin{algorithm}
\caption{: Compute Orthogonal Voronoi Treemap (single layer)}
\label{alg:treemap}
\textbf{Input:} $S$; $\Omega$; $W$; $i_{max}$; $E_{threshold}$;\\
\textbf{Output:} $\upsilon(S)$; \\
\textit{ComputeOVTreemap($S$, $\Omega$, $W$)}
\begin{algorithmic}[1]
%\STATE Initialize $\upsilon(S) = [\ ]$ ;
\STATE \textit{Initialization}($S$, $\Omega$, $W$) ;
\STATE $error = +\infty$ ;
\STATE $\upsilon(S) =  $\textit{ComputeWOVDiagram}($S$, $\Omega$, $W$)  ;
\STATE \textbf{for} $i = 1:i_{max}$
\STATE \qquad $[S, W]$ = \textit{AdaptPositionsWeights}($S$, $\upsilon(S)$, $W$) ;
\STATE \qquad $\upsilon(S) =  $\textit{ComputeWOVDiagram}($S$, $\Omega$, $W$)  ;
\STATE \qquad $error = \dfrac {\sum _{s_{i}\in S}\left| A\left( \upsilon\left( s_{i}\right) \right) -A\left( \Omega \right) * \dfrac {value_{s_{i}}}{value_{S} }\right| }{A\left( \Omega \right) }$ ;
\STATE \qquad \textbf{if} $error < E_{threshold}$, \textbf{then}
\STATE \qquad \qquad \textbf{return} $\upsilon(S)$ ;
\STATE \qquad \textbf{end}
\STATE \textbf{end}
\STATE \textbf{return} $\upsilon(S)$ ;
\end{algorithmic}
\end{algorithm}

An initialization process is first conducted on the hierarchical data to generate the initial positions and weight for each site. This is achieved by the function \textit{Initialization}($S$, $\Omega$, $W$) (Line 1). Then, an initial weighted centroidal orthogonal Voronoi diagram by the function \textit{ComputeWOVDiagram}($S$, $\Omega$, $W$) is generated based on the initial status of sites (Line 3). After that, an iterative process is taken to update the site status by the function \textit{AdaptPositionsWeights}($S$, $\upsilon(S)$, $W$) and redraw the diagram until meeting the converge requirement. To handle hierarchical data with multiple layers, then the function \textit{ComputeOVTreemap($S$, $\Omega$, $W$)} will be called recursively.

\subsection{Initialization}
\label{subsec:initialization}
The initialization step refers to the determination of the initial position of each site and the initial value of weight which is used to control the area of cell. Instead of a random initial value, a reasonable initial site status can significantly improve the algorithm performance. In this section, we determine the layout initial status including the initial status and the initial weight with the help of the treemap layout.

\begin{figure}[tb]
    \centering
    \includegraphics[width= 0.6\columnwidth]{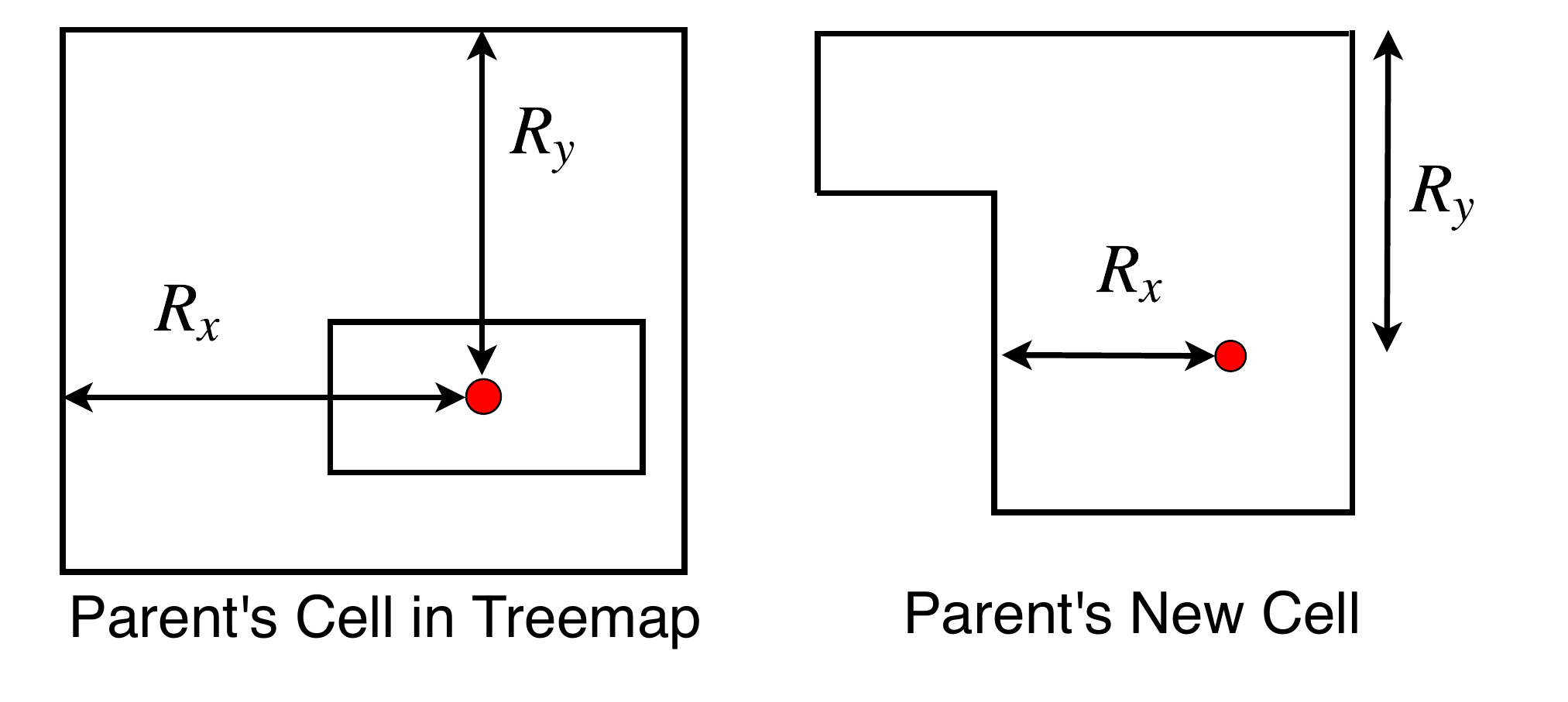}%   
    \caption{Transform the position of a site in rectangle cell of the treemap into the orthogonal rectangle cell as the initial position.}%
    \label{fig:initial}%
\end{figure}

To start with, the hierarchical dataset is visualized by the squarified treemap algorithm~\cite{Bruls:00} since it has been proved that the squarified treemap has a good aspect ratio~\cite{Sondag:18}. It should be noted other treemap algorithm also can be used here. Once the treemap is plotted, we generate a site at the center of each rectangle and calculate the relative position of this site respect to its parent's rectangle boundary. As shown in Fig.~\ref{fig:initial} (left), the value of $R_x$ and $R_y$ is then rescaled to the interval $[0,1]$, respecting to the size of the outer rectangle. This relative position is set as the initial position of this site. When it is used in our algorithm, this relative position of that site is decoded according to its parent's new cell. This decoding is conducted by considering $R_y$ first and then $R_x$. In this way, the position of sites in the treemap will be transformed into the orthogonal rectangles of our plot as the initial position. For the initial weight value, we set it to the half of the area of the site's cell in the treemap. Our experiments show that with these initial setting the final layout is much better than that with random initial status.

%Add Figure Here

\subsection{Site Status Update}
\label{subsec:update}

To update the weight value and the position of each site, we make modification on the strategy used by Hahn et al.~\cite{Hahn:14} since our orthogonal segmentation is different. Instead of updating position and weight separately, we merge these two functionalities into one and make modifications based on our case. The pseudo code for the function \textit{AdaptPositionsWeights($S$, $\upsilon(S)$, $W$)} is depicted in Algorithm~\ref{alg:adapt}. The main difference between our new \textit{AdaptPositionsWeights($S$, $\upsilon(S)$, $W$)} and previous methods~\cite{Nocaj:12, Hahn:14} is that we do not need to check the maximum weight value allowed for each site during the adjustment. The reason is that when we generate the segmentation lines, we have taken the overweight case into consideration (Sect.~\ref{subsubsec:dist}) in order to avoid the generation of empty cells.

\begin{algorithm}
\caption{: Adapt positions and weights.}
\label{alg:adapt}
\textbf{Input:} $S$; $\upsilon(S)$; $W$;  \\
\textbf{Output:} $S$; $W$; \\
\textit{AdaptPositionsWeights($S$, $\upsilon(S)$, $W$)}
\begin{algorithmic}[1]
\STATE $f_s = 0$ ;
\STATE \textbf{foreach} $s_i \in S $
\STATE \qquad $c_i = centroid(\upsilon(s_i))$ ;
\STATE \qquad $A_{current} = A(\upsilon(s_i))$ ;
\STATE \qquad $A_{target} = A(\Omega) * \dfrac{value_{s_i}}{value_{S}}$ ;
\STATE \qquad $f_{adapt} = \dfrac{A_{target}}{A_{current}}$ ;
\STATE \qquad \textbf{if} $f_s \neq 0$ \textbf{and} $sgn(f_{adapt}-1) \neq sgn(f_s-1)$, \textbf{then} 
\STATE \qquad \qquad $f_{adapt} = min(1+\rho, max(f_{adapt}, 1-\rho))$ ;
\STATE \qquad \textbf{end}
\STATE \qquad $w_i = max(w_i * \sqrt{f_{adapt}}, \  \epsilon)$ ;
\STATE \qquad $f_s = f_{adapt}$ ;
\STATE \qquad $s_i^* = s_i + (c_i-s_i)*(1-0.5\rho)$ ;
\STATE \qquad \textbf{if} $s_i^*$ within $\upsilon(s_i)$, \textbf{then} 
\STATE \qquad \qquad $s_i = s_i^*$ ;
\STATE \qquad \textbf{else} 
\STATE \qquad \qquad $s_i = s_i$ ;
\STATE \qquad \textbf{end}
\STATE \textbf{end}
\STATE \textbf{return} $S$, $W$;
%\STATE
%\STATE \qquad 
%\STATE \qquad \qquad 
%\STATE \qquad \qquad \qquad 
\end{algorithmic}
\end{algorithm}

\subsection{Computation of the Orthogonal Voronoi Diagram}
\label{subsec:diagram}
In this section, we describe the computation of the proposed orthogonal Voronoi diagram. Since the diagram is formed by segmentation lines between sites, the strategy to partition the space between two sites is important. According to Eq.~\ref{eq:cell}, the space partitioning can be treated as a clustering of points based on the sites, if the whole plane is covered by uniformly distributed points. There are two main problems arose here. The first problem is how to calculate the distance to the sites, while the second is that for a certain point which site should it belong to. These two problems are straightforward when using the Euclidean distance in the Voronoi diagram. However, they are tough when trying to axis-alignedly subdivide the bounded region. To achieve this, we introduce a new distance function in Sect.~\ref{subsubsec:dist} as well as how to find the reference site among multiple sites in Sect.~\ref{subsubsec:neighbor}. 

Once the partition between two sites is confirmed, how to allocate the partitioned spaces to form a diagram is considered. Here, a sweepline + skyline heuristic algorithm is introduced in Sect.~\ref{subsubsec:sweepline}.

\subsubsection{Distance Function}
\label{subsubsec:dist}
%Add Figure Here
\begin{figure}[tb]
    \centering
    \includegraphics[width = 0.6 \columnwidth]{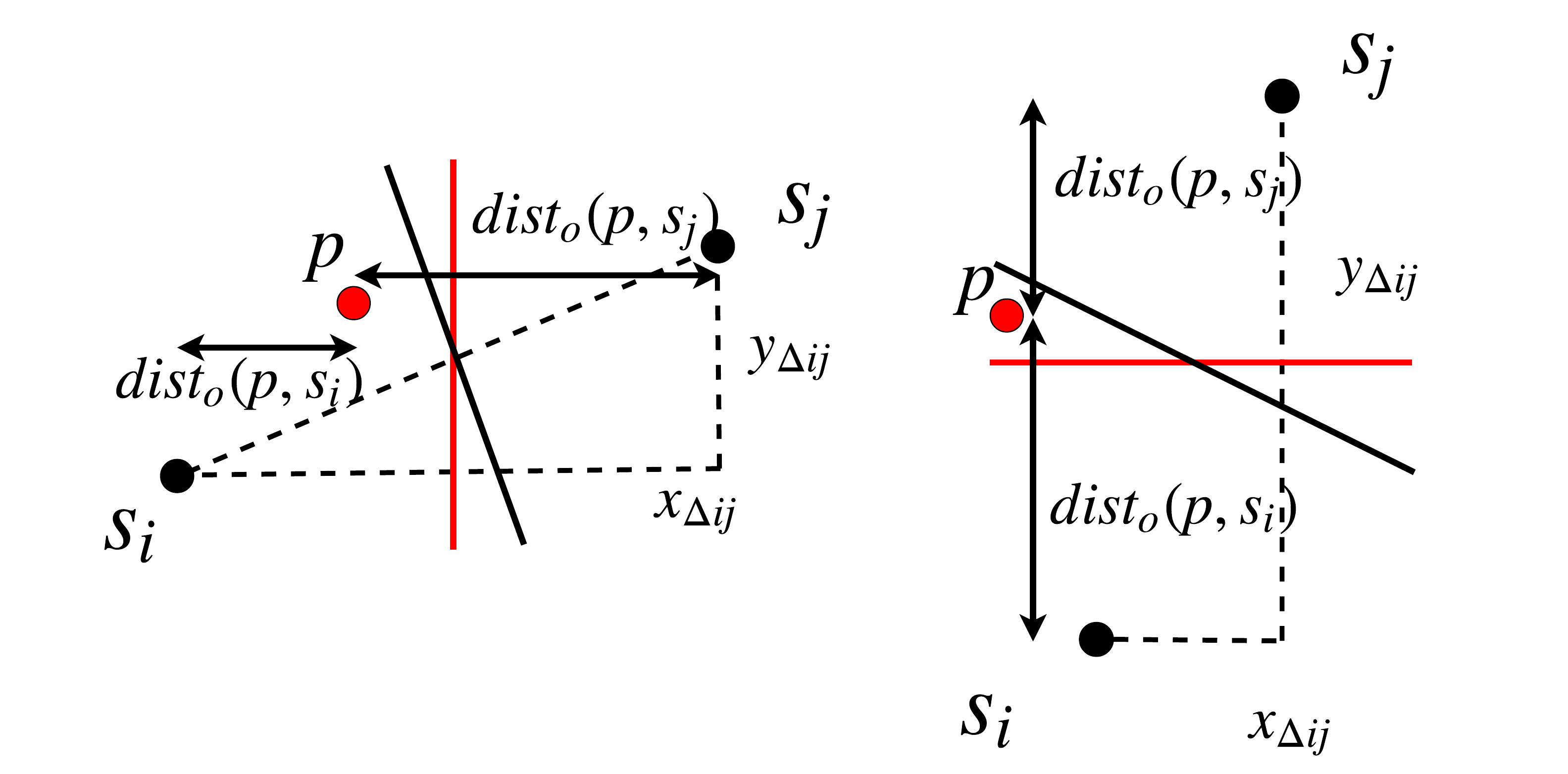}  
    \caption{ Two kinds of relative positions for sites $s_i$ and $s_j$ where $x_{\Delta ij} = | x_i - x_j|$ and $y_{\Delta ij} = | y_i - y_j |$. The black solid line is the segmentation based on Euclidean distance while the red solid line is an axis-aligned segmentation. The distance $dist_{o}(p, s_i)$ and $dist_{o}(p, s_j)$ in both cases are illustrated.}%
    \label{fig:dist1}%
\end{figure}

The new distance function considers the relative positions of two sites rather than only one site in previous distance function. Formally, when calculating the distance of point $p$ and site $s_i$, we consider the relative positions of site pair $s_i$ and $s_j$ to decide which coordinate should be considered. Since we try to divide the space axis-aligned, we consider two kinds of relative positions between site $s_i$ and $s_j$ based on their x-axis difference $x_{\Delta ij} = | x_i - x_j|$ and y-axis difference $y_{\Delta ij} = | y_i - y_j |$. Then for an arbitrary point $p$, the distance to site $s_i$ in both cases are depicted in Fig.~\ref{fig:dist1} and defined as:
\begin{equation}
\label{eq:o}
dist_{o}(p, s_i) = 
\begin{cases}
| x_p - x_{s_i} |  \qquad if \ x_{\Delta ij} > y_{\Delta ij}, \\ 
| y_p - y_{s_i} |  \qquad if \ x_{\Delta ij} < y_{\Delta ij}.
\end{cases}
\end{equation}

When a positive weight value is associated with each site, the distance function (Eq.~\ref{eq:o}) should be modified as:
\begin{equation}
\label{eq:wo}
dist_{o}(p, s_i) = 
\begin{cases}
| x_p - x_{s_i} | - w_i \qquad if \ x_{\Delta ij} > y_{\Delta ij}, \\ 
| y_p - y_{s_i} | - w_i \qquad if \ x_{\Delta ij} < y_{\Delta ij}.
\end{cases}
\end{equation}

For the left-hand-side case in Fig.~\ref{fig:dist1}, a vertical line is needed to separate the site pair while a horizontal line for the right-hand-side case in Fig.~\ref{fig:dist1}. Hence, the axis-aligned segmentation line $L$ is defined as:
\begin{equation}
\label{eq:line1}
L_{s_{i}s_{j}}:
\begin{cases}
x=\dfrac {1}{2}\left( x_{s_i}+x_{s_j}\right) \qquad if \ x_{\Delta ij} > y_{\Delta ij}, \\ 
y=\dfrac {1}{2}\left( y_{s_i}+y_{s_j}\right) \qquad if \ x_{\Delta ij} < y_{\Delta ij}.
\end{cases}
\end{equation}
If two sites have large $y_{\Delta ij}$, then they will be separated by a horizontal line. Otherwise, these two sites will be separated by a vertical line. For the case that $x_{\Delta ij} = y_{\Delta ij}$, we break the tie by choosing a horizontal segmentation. 

To guarantee that there is no cell with empty region, the segmentation lines should be located in between. This is based on an assumption that the sum of $w_i$ and $w_j$ are smaller than the distance between two sites. However, it is possible that this assumption fails. In this case, we partition the space according to the ratio of the weight values by the function \textit{GenerateWLine($s_i$, $s_j$, $w_i$, $w_j$)}. A pseudo-code for this function is depicted in Algorithm~\ref{alg:line}. 

\begin{algorithm}
\caption{: Generate Weighted Segmentation Line}
\label{alg:line}
\textbf{Input:} $s_i$; $s_j$; $w_i$; $w_j$; \\
\textbf{Output:} $L_{s_{i}s_{j}}$; \\
\textit{GenerateWLine($s_i$, $s_j$, $w_i$, $w_j$)}
\begin{algorithmic}[1]
\STATE \textbf{if} site $s_i$ and $s_j$ are horizontal neighbors, \textbf{then}
\STATE \qquad \textbf{if} $|x_{s_i}-x_{s_j}| - w_i - w_j >= 0$, \textbf{then} $L_{s_{i}s_{j}} = min(x_{s_i}+w_i, x_{s_j}+w_j) + \dfrac {1}{2}\left( |x_{s_i}-x_{s_j}| - w_i - w_j \right)$;
\STATE \qquad \textbf{else} $L_{s_{i}s_{j}} = (x_{s_i}<x_{s_j})\ ? \ x_{s_i} + \dfrac {w_i}{w_i+w_j} * |x_{s_i}-x_{s_j}| : x_{s_j} + \dfrac {w_j}{w_i+w_j} * |x_{s_i}-x_{s_j}|$; 
\STATE \qquad \textbf{end}
\STATE \textbf{else}
\STATE \qquad \textbf{if} $|y_{s_i}-y_{s_j}| - w_i - w_j >= 0$, \textbf{then} $L_{s_{i}s_{j}} = min(y_{s_i}+w_i, y_{s_j}+w_j) + \dfrac {1}{2}\left( |y_{s_i}-y_{s_j}| - w_i - w_j \right)$;
\STATE \qquad \textbf{else} $L_{s_{i}s_{j}} = (y_{s_i}<y_{s_j})\ ?\ y_{s_i} + \dfrac {w_i}{w_i+w_j} * |y_{s_i}-y_{s_j}| : y_{s_j} + \dfrac {w_j}{w_i+w_j} * |y_{s_i}-y_{s_j}|$; 
\STATE \qquad \textbf{end}
\STATE \textbf{end}
\STATE \textbf{return} $L_{s_{i}s_{j}}$;
\end{algorithmic}
\end{algorithm}

\subsubsection{Clustering}
\label{subsubsec:neighbor}
%Add Figure Here
\begin{figure}[tb]
    \centering
    \subfloat[]{{\includegraphics[width= 0.4 \columnwidth]{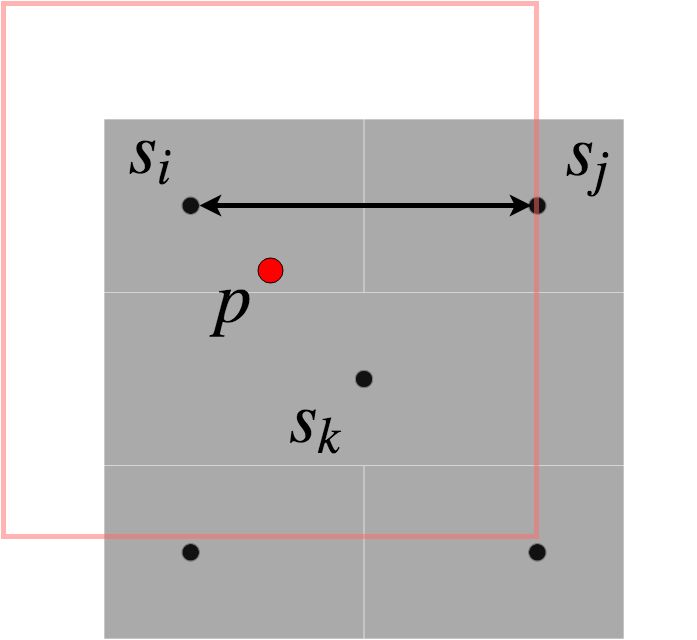} }}%   
    \caption{One example for choosing the reference sites for certain point $p$ in red. A square in pink centered at the point $p$ is built, including at least one horizontal site pair and one vertical site pair. By first considering horizontal site pair $s_i$ and $s_j$, and then vertical site pair $s_i$ and $s_k$, point $p$ has smallest distance to site $s_i$.}%
    \label{fig:dist2}%
\end{figure}

Once the distance function is confirmed, the second question is that for point $p$, which site should it belong to after calculating the distance. According to Eq.~\ref{eq:wo}, point $p$ may have multiple distance value to the same site when considering different site pairs. To eliminate the confusion, we first define the concept of valid neighborhood relationship between sites. Then we describe a principle to handle the case with multiple site pairs from the perspective of point $p$. 

For a site pair, if there are no other sites located in the rectangle region formed by the two sites, then these two sites are valid neighbors. In terms of the relative positions, these two sites are either horizontal neighbors (i.e. neighbors in the horizontal direction) or vertical neighbors (i.e. neighbors in the vertical direction). The valid neighbor sites can also be considered from the perspective of the final diagram. If the existence of site $s_j$ contributes to the generation of the cell of site $s_i$, then site $s_i$ and $s_j$ are valid neighbors.

For the case when there are multiple sites around, we define a principle to do clustering. For easy understanding, we take the case in Fig.~\ref{fig:dist2} as an example. Firstly, we design the smallest square (in pink) centered at point $p$, including multiple sites such that there are at least one pair of vertical valid neighbor sites ($s_i$ and $s_k$) and one pair of horizontal valid neighbor sites ($s_i$ and $s_j$). Secondly, the segmentation line between the horizontal neighbor sites determines which site the point $p$ should belong to. Thirdly, suppose point $p$ belongs to site $s_i$ now, then check the distance between point $p$ and all of the vertical neighbors of site $s_i$ to find the smallest one. If site $s_i$ has the smallest or second smallest vertical distance to point $p$, then point $p$ will be clustered accordingly. Otherwise, extend the square until such a kind of case meets. 

However, this principle is complex and not automatic. Hence, a computer-based algorithm should be designed to automatically execute this principle. In the next section, a sweepline + skyline algorithm is described to carry out the partitioning based on the new distance function.

\subsubsection{Sweepline + skyline algorithm}
\label{subsubsec:sweepline}

\begin{algorithm}
\caption{: Compute Orthogonal Voronoi Diagram}
\label{alg:diagram}
\textbf{Input:} $S$; $\Omega$; $W$; \\
\textbf{Output:} $\upsilon(S)$; \\
\textit{ComputeWOVDiagram ($S$, $\Omega$, $W$)}
\begin{algorithmic}[1]
%\STATE Initialize the position of $S$ by random ;
\STATE Sort $S$ in the order of $x_{s_i}$ ascending ;
\STATE Initialize $L_{sweepline}$ ;
\STATE Initialize $L_{skyline}$ ;
\STATE Initialize $\upsilon(S) = [\ ]$ ;
\STATE \textbf{for} $i = 1:n$
\STATE \qquad \textbf{for} $j = 1:i-1$
\STATE \qquad \qquad \textbf{if} site $s_j$ is closed, \textbf{then} 
\STATE \qquad \qquad \qquad continue;
\STATE \qquad \qquad \textbf{end}
\STATE \qquad \qquad \textbf{if} site $s_j$ is not the valid neighbor of site $s_i$, \textbf{then} 
\STATE \qquad \qquad \qquad continue;
\STATE \qquad \qquad \textbf{end}
\STATE \qquad \qquad \textbf{if} site $s_i$ and $s_j$ are vertical neighbors, \textbf{then}
\STATE \qquad \qquad \qquad $L_{s_{i}s_{j}}$ = \textit{GenerateLine($s_i$,$s_j$, $w_i$, $w_j$)} ;
\STATE \qquad \qquad \qquad Update $L_{skyline}$ with $L_{s_{i}s_{j}}$;
\STATE \qquad \qquad \textbf{else}
\STATE \qquad \qquad \qquad $L_{s_{i}s_{j}}$ = \textit{GenerateLine($s_i$,$s_j$, $w_i$, $w_j$)} ;
\STATE \qquad \qquad \qquad Mark site $s_j$ as closed;
\STATE \qquad \qquad \qquad Generate the bounding polygon $\upsilon(s_j)$ for site $s_j$;
\STATE \qquad \qquad \qquad $\upsilon(S).push(\upsilon(s_j))$;
\STATE \qquad \qquad \qquad Update $L_{skyline}$ with $L_{s_{i}s_{j}}$;
\STATE \qquad \qquad \textbf{end}
\STATE \qquad \textbf{end}
\STATE \qquad Update $L_{sweepline}$;
\STATE \textbf{end}
\STATE \textbf{for} $i = 1:n$
\STATE \qquad \textbf{if} site $s_i$ is closed, \textbf{then} continue;
\STATE \qquad \textbf{else} 
\STATE \qquad \qquad Generate the bounding polygon $\upsilon(s_i)$ for site $s_i$;
\STATE \qquad \qquad $\upsilon(S).push(\upsilon(s))$;
\STATE \qquad \textbf{end}
\STATE \textbf{end}
\STATE \textbf{return} $\upsilon(S)$;
\end{algorithmic}
\end{algorithm}

%Add Figure Here
\begin{figure}[tb]
    \centering
    \subfloat[]{{\includegraphics[width= 0.31 \columnwidth]{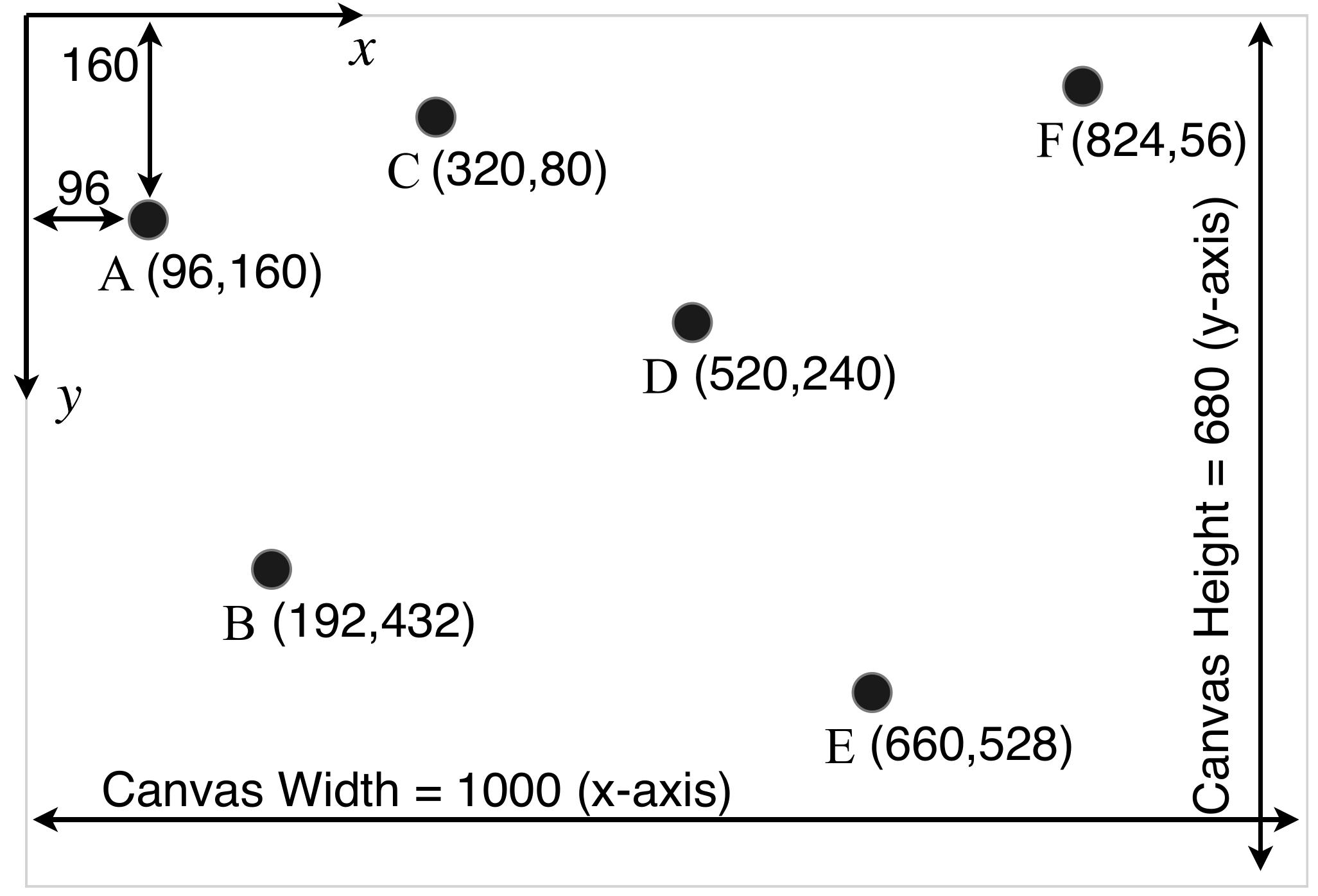} }}%
    \quad
    \subfloat[]{{\includegraphics[width= 0.31 \columnwidth]{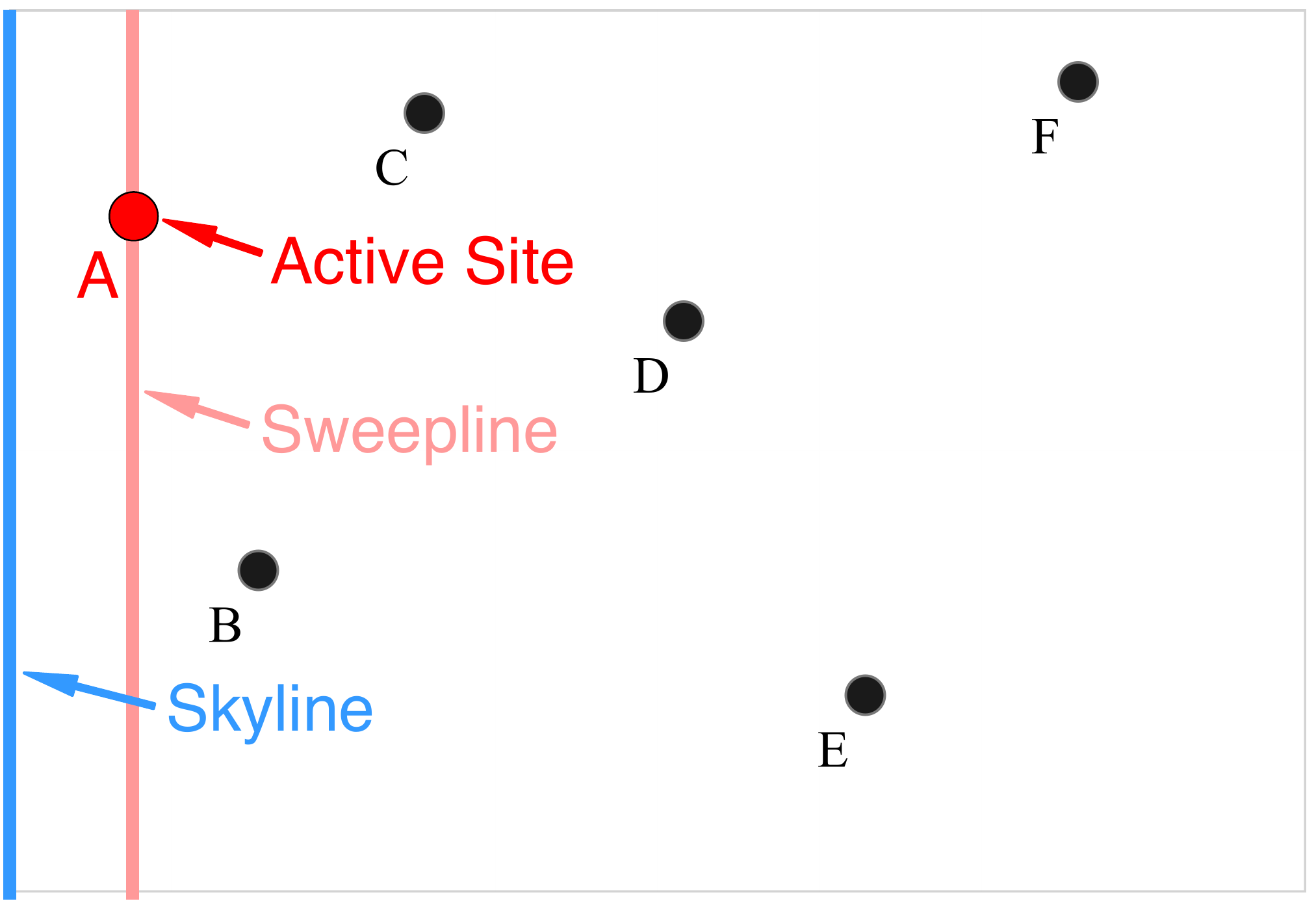} }}%
    \quad
    \subfloat[]{{\includegraphics[width= 0.31 \columnwidth]{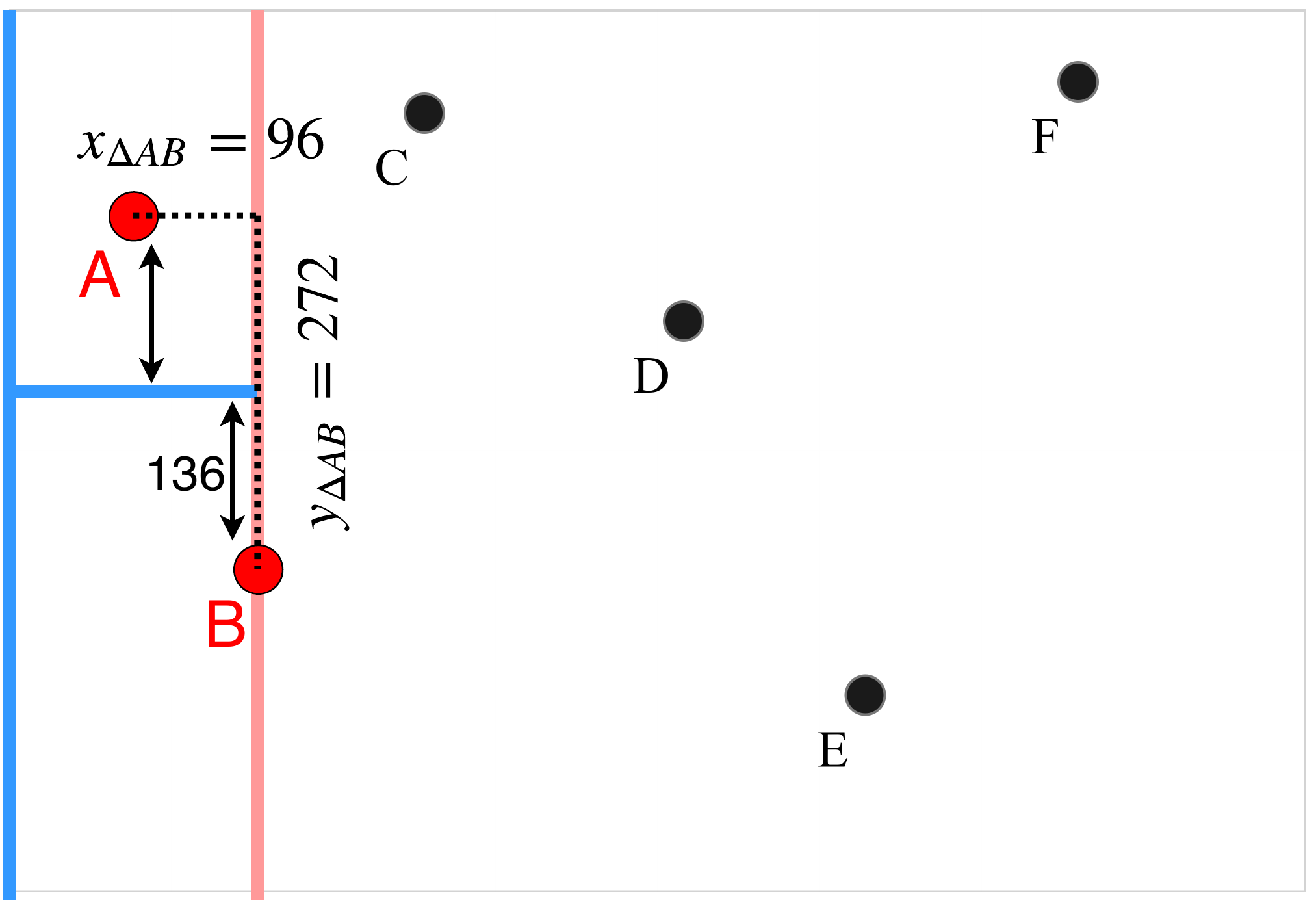} }}%
    \\
    \subfloat[]{{\includegraphics[width= 0.31 \columnwidth]{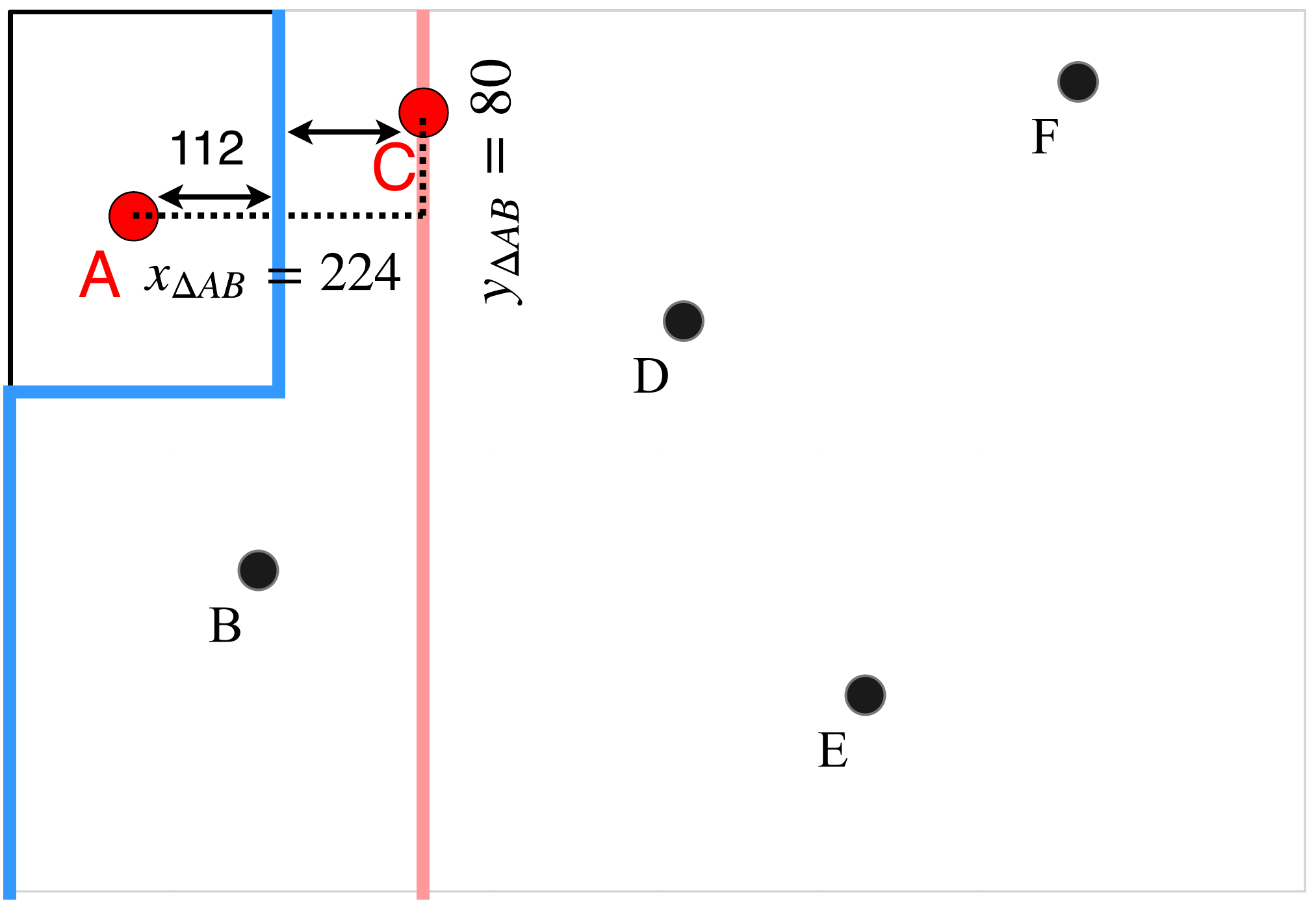} }}%
    \quad
    \subfloat[]{{\includegraphics[width= 0.31 \columnwidth]{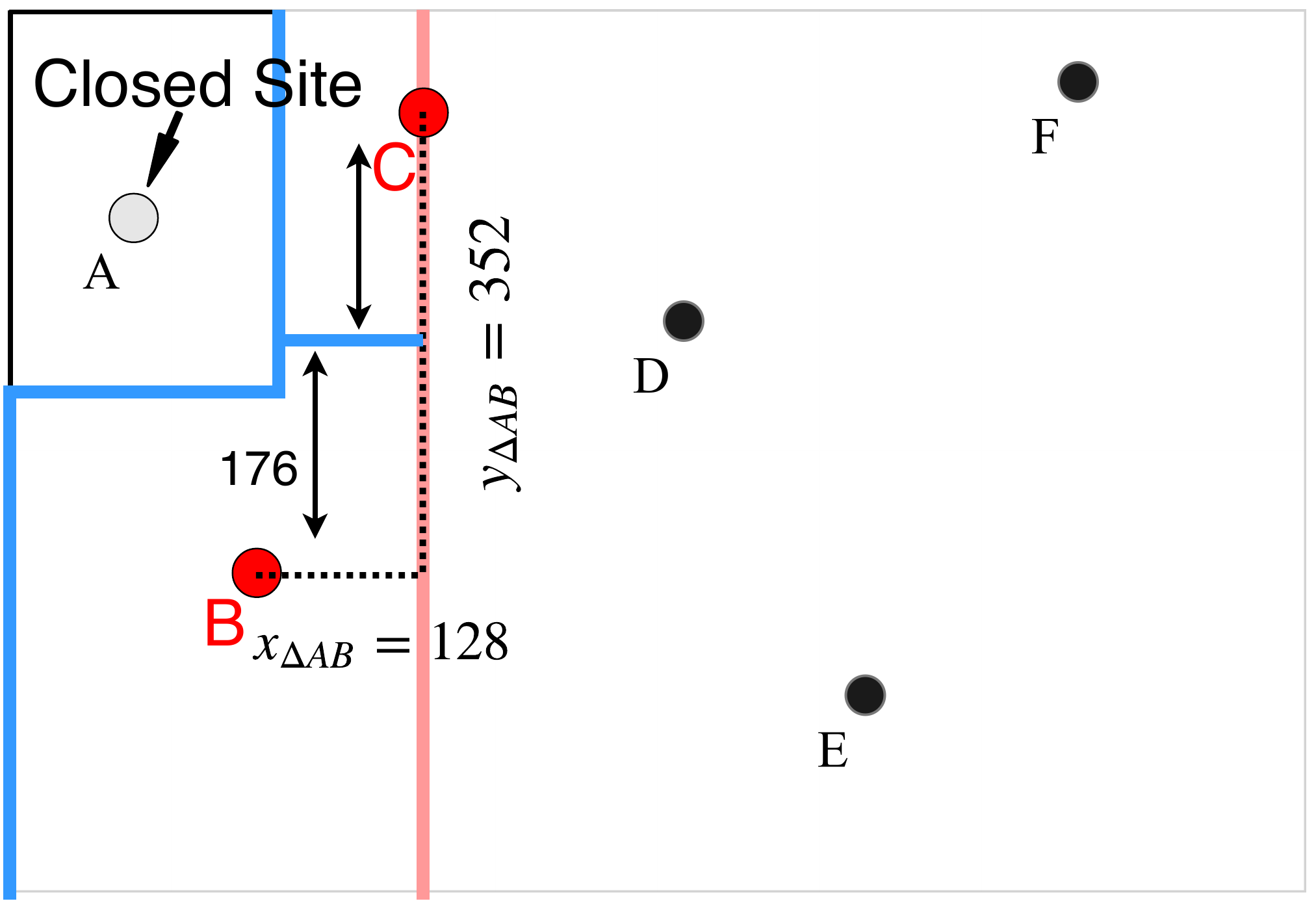} }}% 
    \quad
    \subfloat[]{{\includegraphics[width= 0.31 \columnwidth]{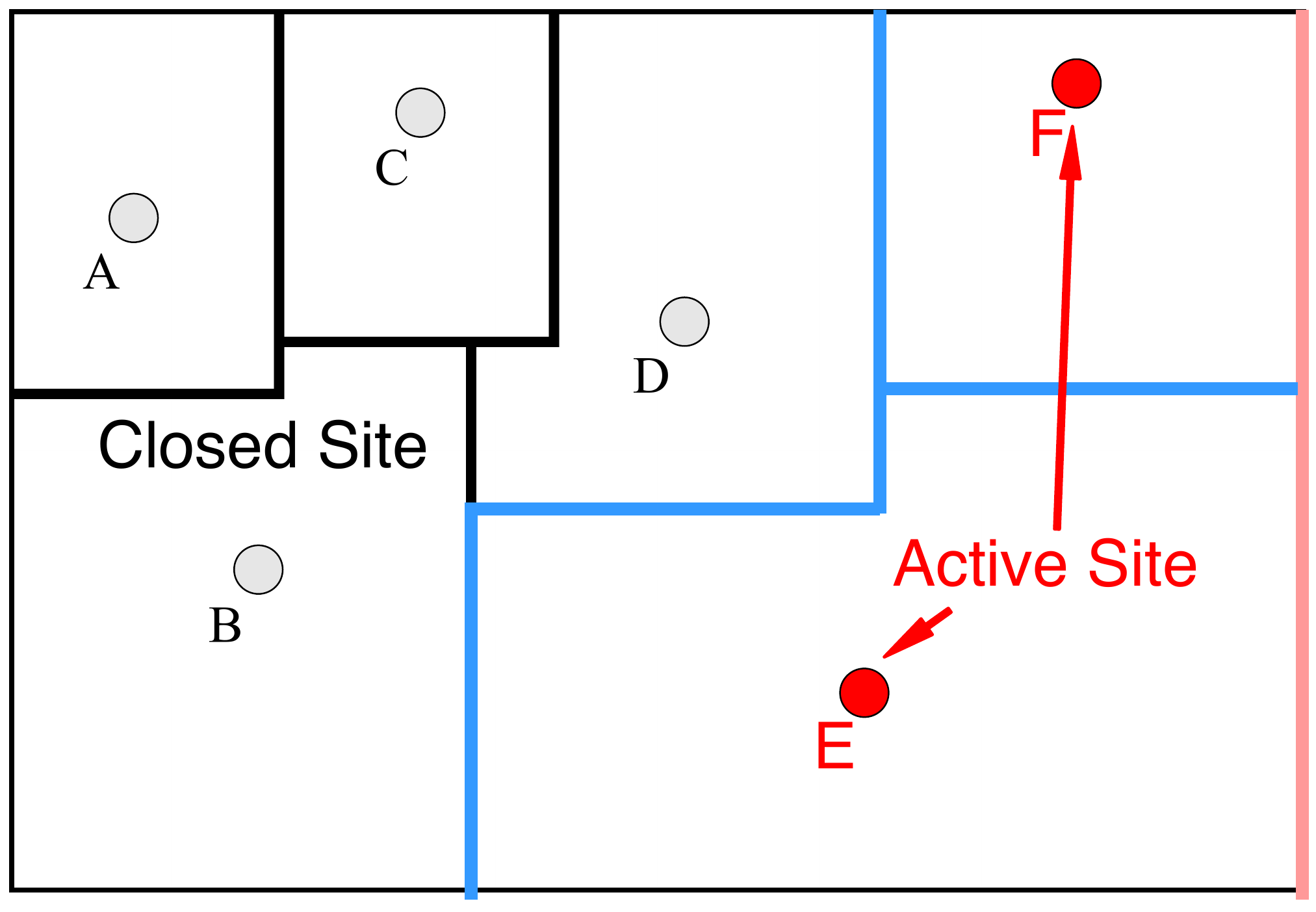} }}%   
    \caption{Overview of the sweepline + skyline algorithm. (a) Initially, six sites are positioned inside the canvas based on their initial positions. (b) The skyline (in blue) and the sweepline (in pink) are initialized and the first site $s_A$ (left-most) is activated. (c) Sweep the sweepline to meet the second site $s_B$ and determine the relationship of $s_A$ and $s_B$. Then a horizontal segmentation line $L_{AB}$ is added to the skyline. (d) Sweep the sweepline to site $s_C$ and determine the relationship of $s_A$ and $s_C$. Since $s_A$ and $s_C$ are horizontal neighbors, a vertical segmentation line is generated. In this case, site $s_A$ is closed. A bounding polygon formed by the skyline and the canvas boundary and the vertical segmentation line is built for $s_A$. Then the skyline is updated and site $s_A$ is marked as closed status. (e) Site $s_B$ and $s_C$ are then discussed and a horizontal segmentation line is added to the skyline. (f) The process is continued until the last site $s_F$ is considered. All the active sites will be closed and the corresponding bounding polygon will be built based on the current skyline and the canvas boundary. }%
    \label{fig:algorithm1}%
\end{figure}

The proposed sweepline + skyline algorithm aims to automatically partition the canvas into orthogonal sub-regions based on the new distance function proposed in Sect.~\ref{subsubsec:dist}. Our idea is motivated by the sweep line algorithm for Voronoi diagram~\cite{Fortune:87} and the skyline strategy used in cutting and packing problem~\cite{Burke:04}. The sweep line used in our algorithm is a vertical line moving from left to right. When the sweep line hits a new site, the relationship of this new site and all its left-hand-side site pairs are checked to generate vertical or horizontal segmentation lines. Meanwhile, a skyline is defined to record the current segmentation lines for all active sites. When the sweep line hits a new site and new segmentation lines are generated, the skyline will be updated correspondingly. 

Figure ~\ref{fig:algorithm1} illustrates an example for this process. As shown in Fig.~\ref{fig:algorithm1} (a), for a given rectangular canvas with $width = 1000$ and $height=680$, six sites are positioned based on their coordinates (we follow the image coordinate system where the y-axis is down). The size of the canvas and the positions of sites are the input value of our algorithm. The first step (Fig.~\ref{fig:algorithm1} (b)) is to create a vertical sweepline and a vertical skyline. The sweepline is initially located on the left-most site $s_A$ while the skyline is on the left-hand-side of the canvas. The length of both lines equal to the height of the canvas. The second step (Fig.~\ref{fig:algorithm1} (c)) is to sweep the sweepline from left to right to hit the next site $s_B$. Since $y_{\Delta AB} = 272$ is larger than $x_{\Delta AB} = 96$, a horizontal line $L_{AB}:{y=296}$ is built between site $s_A$ and $s_B$. The skyline is then updated by adding a horizontal line segment. After checking all the left-hand-side site pairs of site $s_B$, the sweepline moves to the next site $s_C$ (Fig.~\ref{fig:algorithm1} (d)). Since site $s_A$ is not closed and is the horizontal neighbor of site $s_C$, then a vertical line $L_{AC}:{x=208}$ is built. Once a vertical line is generated, the left site $s_A$ of this horizontal neighbor sites should be closed and the bounding polygon of site $s_A$ is formed based on the current skyline and the new vertical segment (as well as the canvas boundary). After that, the skyline is updated. Since site $s_B$ is not closed and is the vertical neighbor of site $s_C$, then a horizontal line $L_{BC}:{y=256}$ is built and the skyline is updated again (Fig.~\ref{fig:algorithm1} (e)). This process is repeated until the sweepline reaches the last site $s_F$ (Fig.~\ref{fig:algorithm1} (f)). The residual sites ($s_E$ and $s_F$) will then be closed and the bounding polygon for each site is formed by the current skyline and sweepline together. A pseudo-code for the whole process is depicted in Algorithm~\ref{alg:diagram}.

\subsection{Implementation}
\label{subsec:implementation}
The implementation of our proposed algorithms is in JavaScript. Our code mainly depends on the D3.js package~\cite{d3} and previous implementation on the Voronoi treemap~\cite{Voronoi}\footnote{ This JavaScript implementation is based on the work by Noca and Brandes~\cite{Nocaj:12} in which the program is run in Java. }. The color schema for some of the plots in this paper is generated by the Tree Color~\cite{Tennekes:14}, such as Fig.~\ref{fig:teaser} (b, c). The source code of our implementation can be found in Github.

\section{Performance}
\label{sec:performance}
\begin{table}[]
\caption{ The running time (in $ms$) for a single iteration.}
\label{tab:time}
\centering%
\begin{tabular}{c|c|c|c|c|c|c|c|c|c|c|c|c}
\hline
                           & 50     & 100    & 150    & 200     & 250     & 300     & 350    & 400    & 450    & 500    & 550    & 600    \\ \hline
VT~\cite{Nocaj:12}           & 0.22   & 0.3    & 0.43   & 0.61    & 0.79    & 1.05    & 1.19   & 1.43   & 1.82   & 1.94   & 2.2    & 2.52   \\ \hline
VT            & 1.38 & 3.96 & 6.83 & 10.95 & 16.59 & 25.13  & 30.17  & 39.48 & 46.30 & 64.45 & 82.17 & 91.34 \\ \hline
OVT & 0.93  & 2.74 & 5.61 & 9.24  & 14.19 & 19.60 & 26.31 & 31.10 & 38.61 & 45.94 & 53.90 & 71.13  \\ \hline
\end{tabular}
\end{table}

%Add Figure Here
\begin{figure}[tb]
    \centering
    \includegraphics[width= 0.6 \columnwidth]{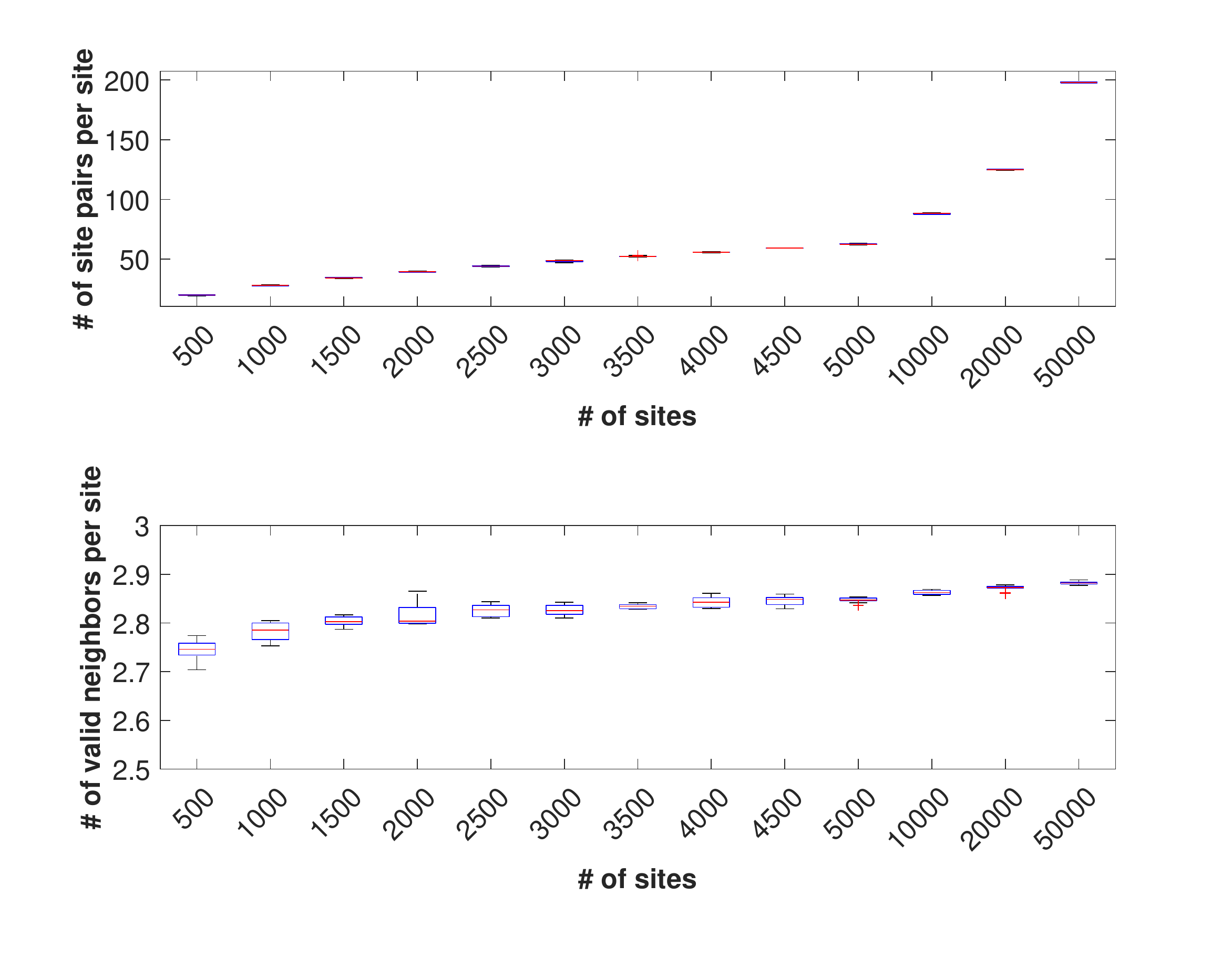} %   
    \caption{The average number of site pairs (top) and valid neighbors (bottom) per site. Noted that the x-axis is not linear.}%
    \label{fig:exp1}%
\end{figure}

%Add Figure Here
\begin{figure}[tb]
    \centering
    \includegraphics[width= 0.6 \columnwidth]{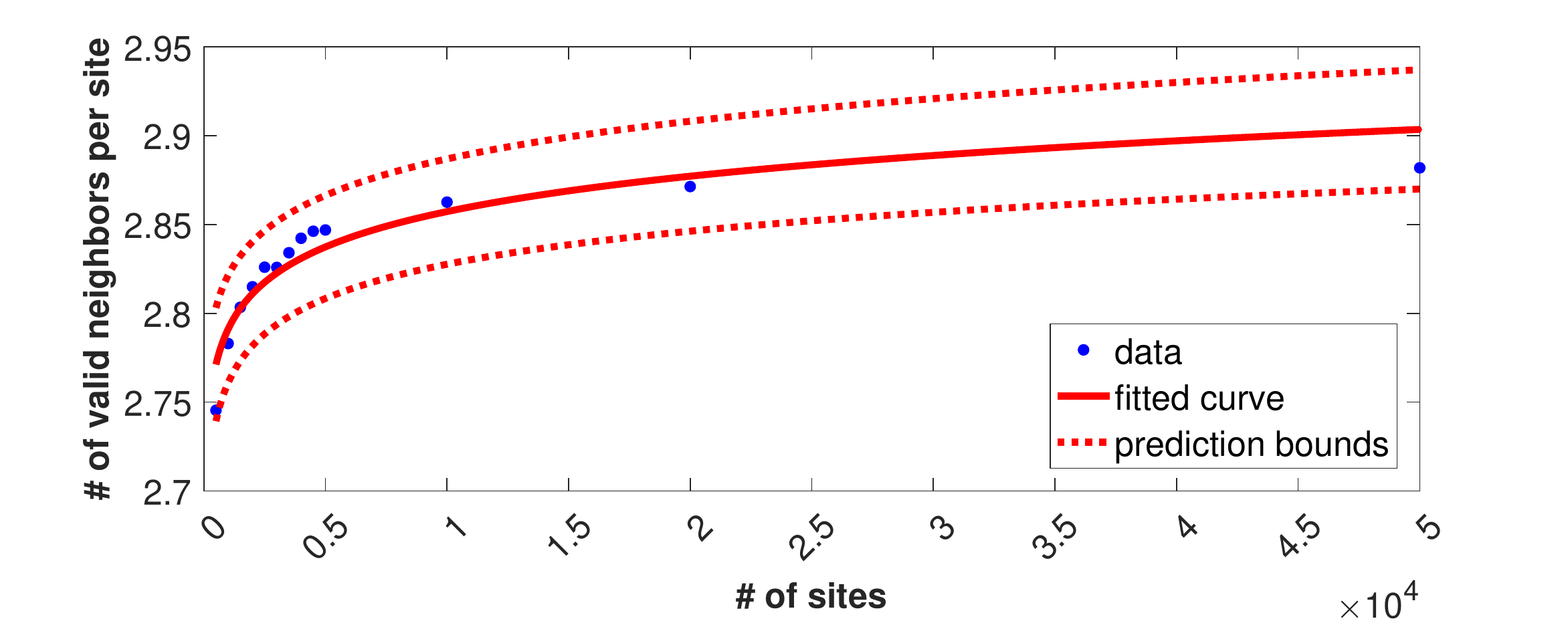} %   
    \caption{The fitting line and the prediction for the average number of valid neighbors per site.}%
    \label{fig:exp1-1}%
\end{figure}

%Add Figure Here
\begin{figure}[tb]
    \centering
    \includegraphics[width= 0.6 \columnwidth]{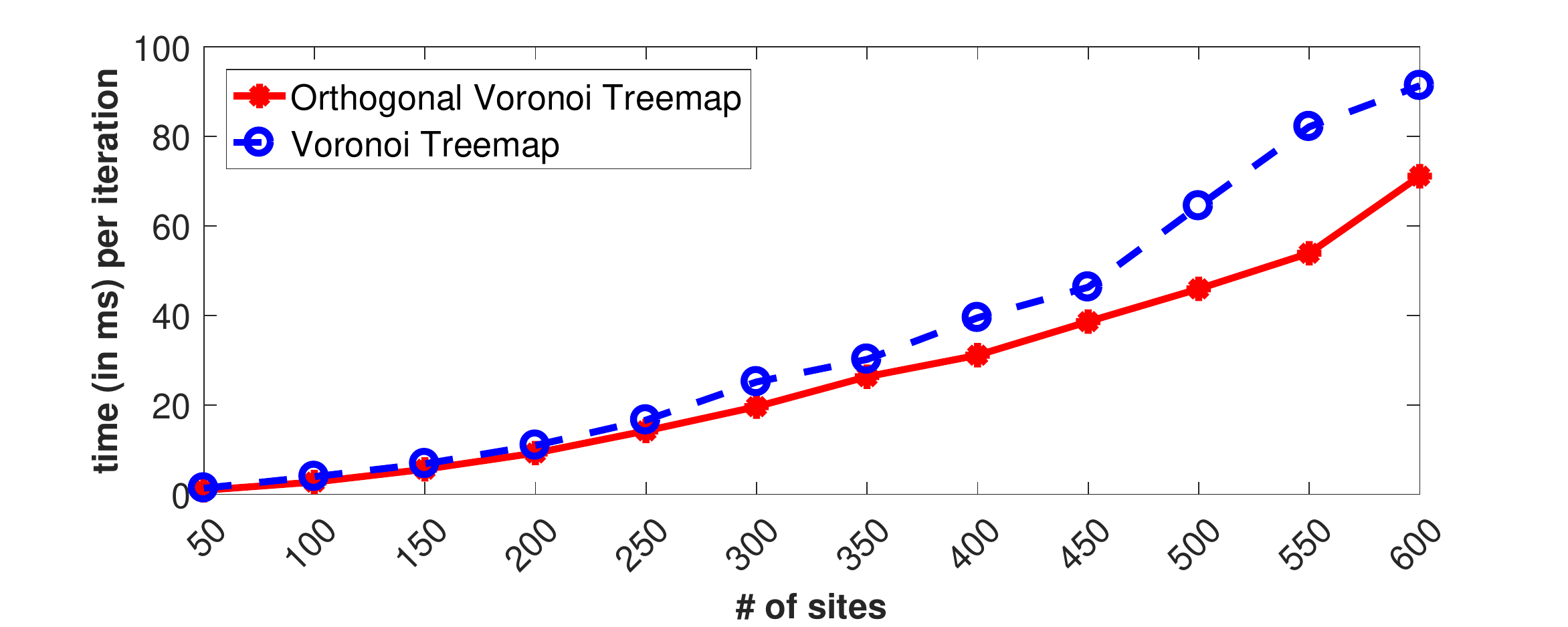} %   
    \caption{The running time (in $ms$) for a single iteration. }%
    \label{fig:exp2}%
\end{figure}

%Add Figure Here
\begin{figure}[tb]
    \centering
    \subfloat[Voronoi Treemap]{{\includegraphics[width=  0.6 \columnwidth]{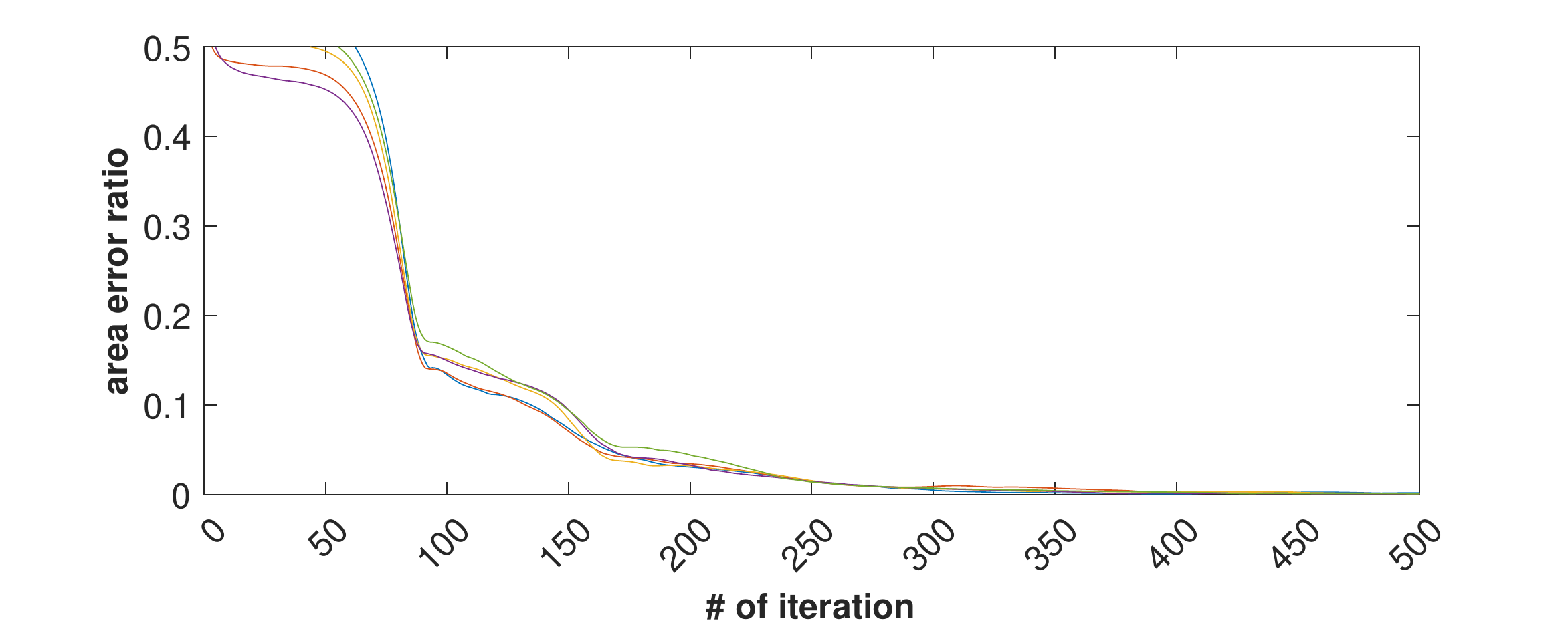} } } %    
    \\    
    \subfloat[Orthogonal Voronoi Treemap with random initial status]{{\includegraphics[width= 0.6  \columnwidth]{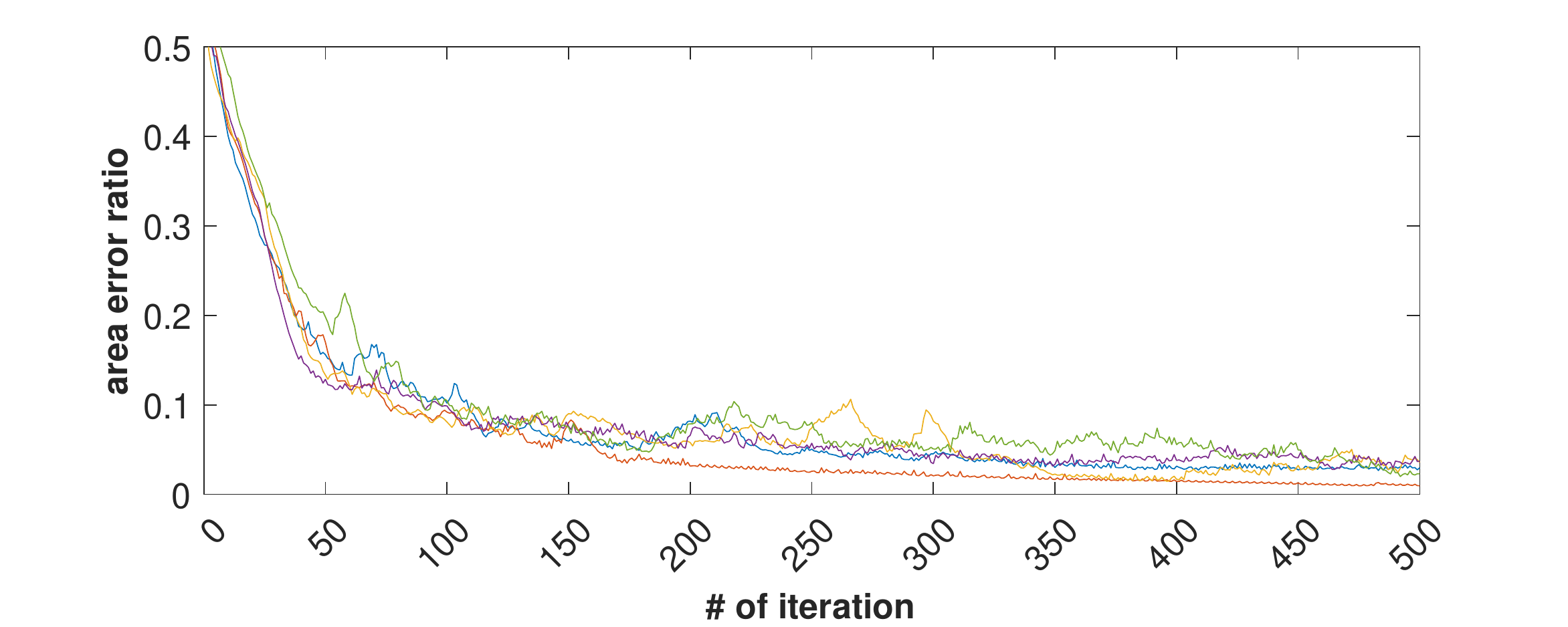} } } %    
    \\    
    \subfloat[Orthogonal Voronoi Treemap with designed initial status]{{\includegraphics[width= 0.6  \columnwidth]{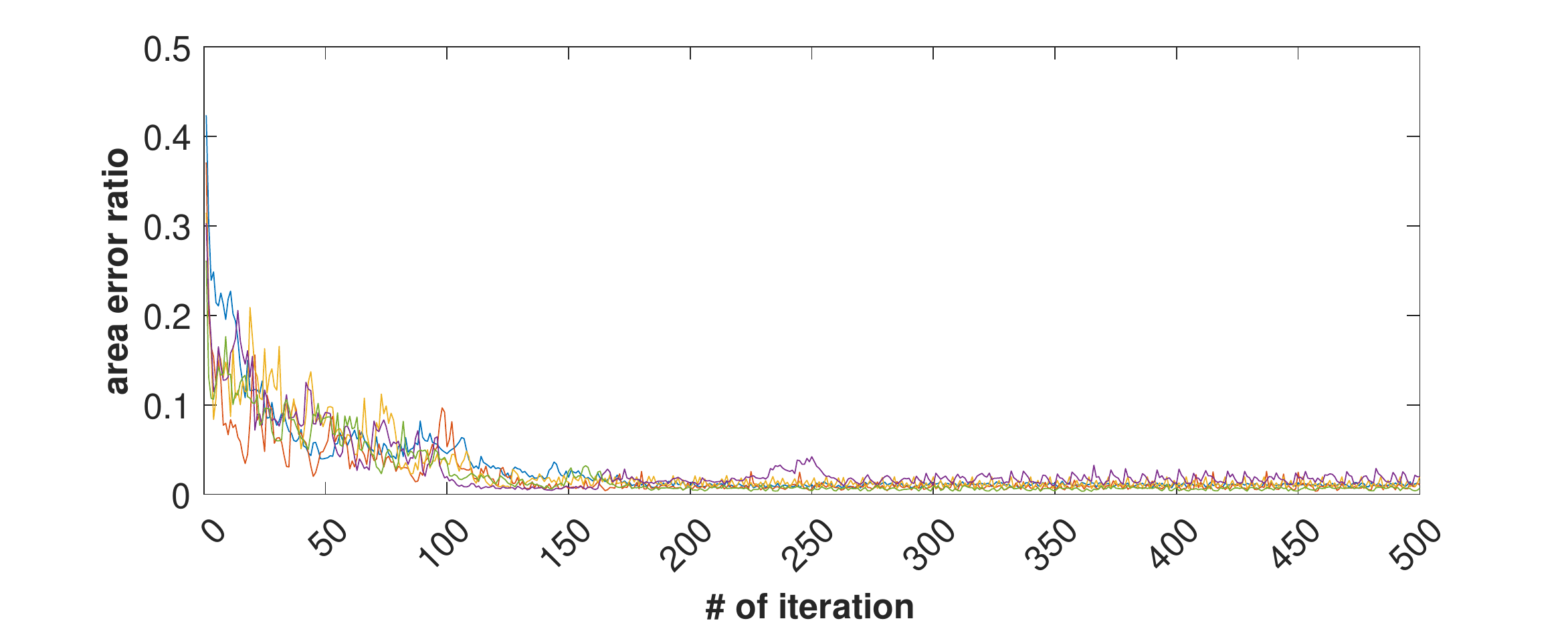} } } %        
    \caption{The converge rate of the Voronoi treemap, our orthogonal Voronoi treemap with random initial status and with our designed initial status. Within the 500 iterations, we record the area error of each iteration to indicate the converge rate. }%
    \label{fig:exp3}%
\end{figure}

%Add Figure Here
\begin{figure}[tb]
    \centering
    \subfloat[Test on random dataset ]{{\includegraphics[width= 0.6 \columnwidth]{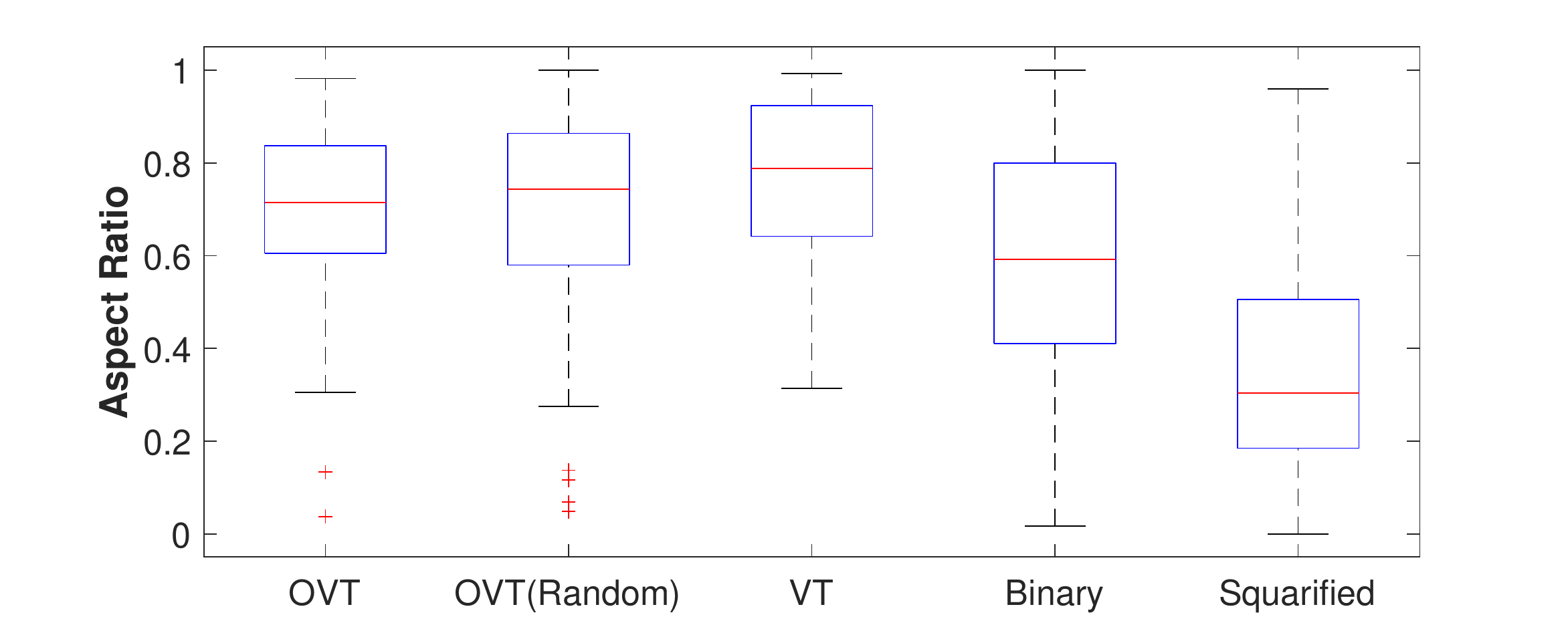} } } %    
    \\
    \subfloat[Test on the globalGDP dataset ]{{\includegraphics[width= 0.6 \columnwidth]{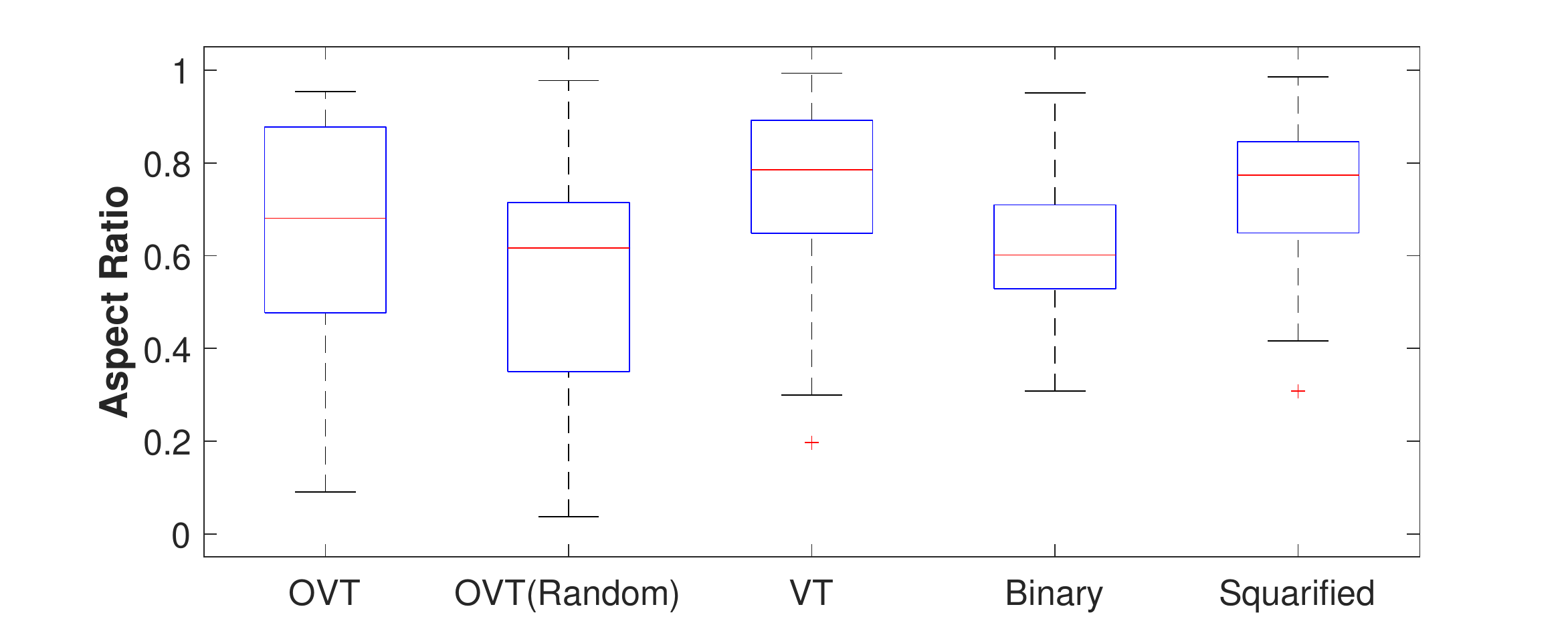} } }%      
    \\
    \subfloat[Test on the Flare dataset ]{{\includegraphics[width= 0.6 \columnwidth]{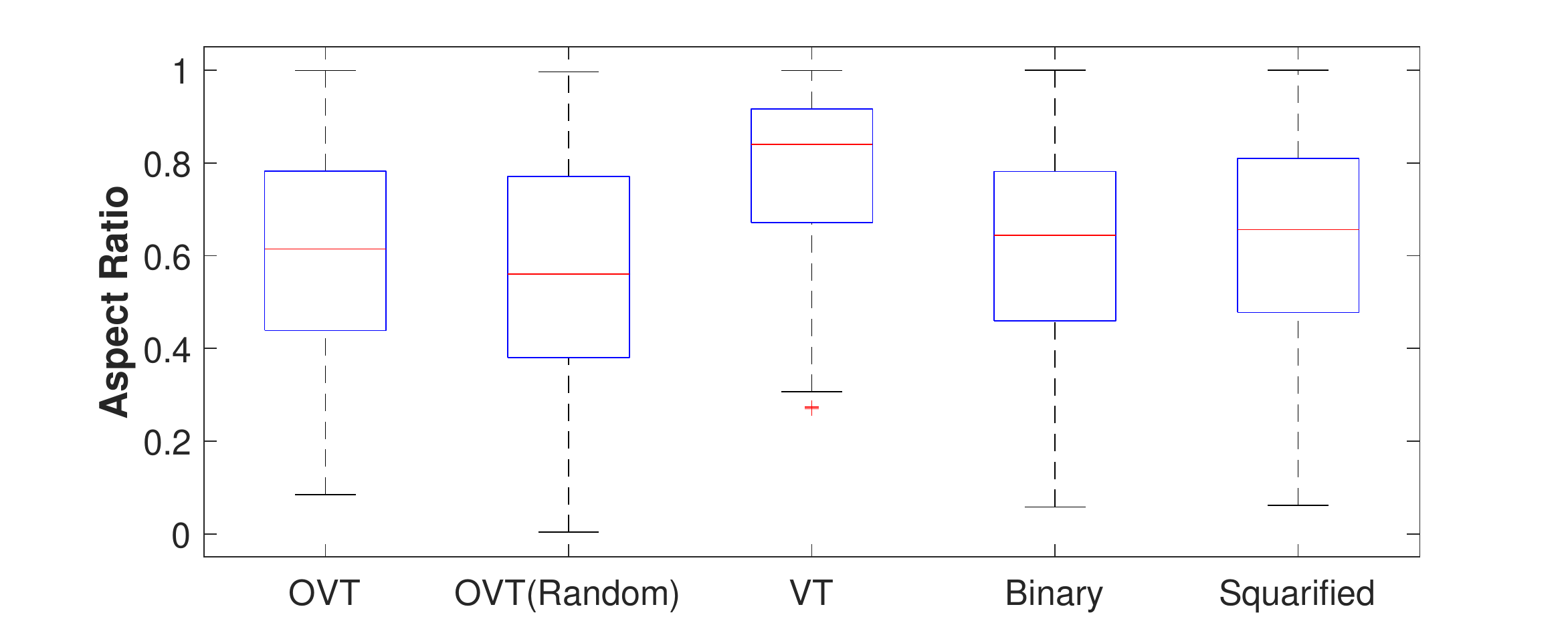} } }%      
    \caption{The boxplots of aspect ratio. In these tests, our orthogonal Voronoi treemap (OVT) has a comparable mean aspect ratio with the treemaps (binary treemap and squarified treemap). While the Voronoi treemap has the best aspect ratio in all tests. }%
    \label{fig:exp4}%
\end{figure}

%Add Figure Here
\begin{figure}[tb]
    \centering
    \subfloat[OVT]{{\includegraphics[width= 0.4 \columnwidth]{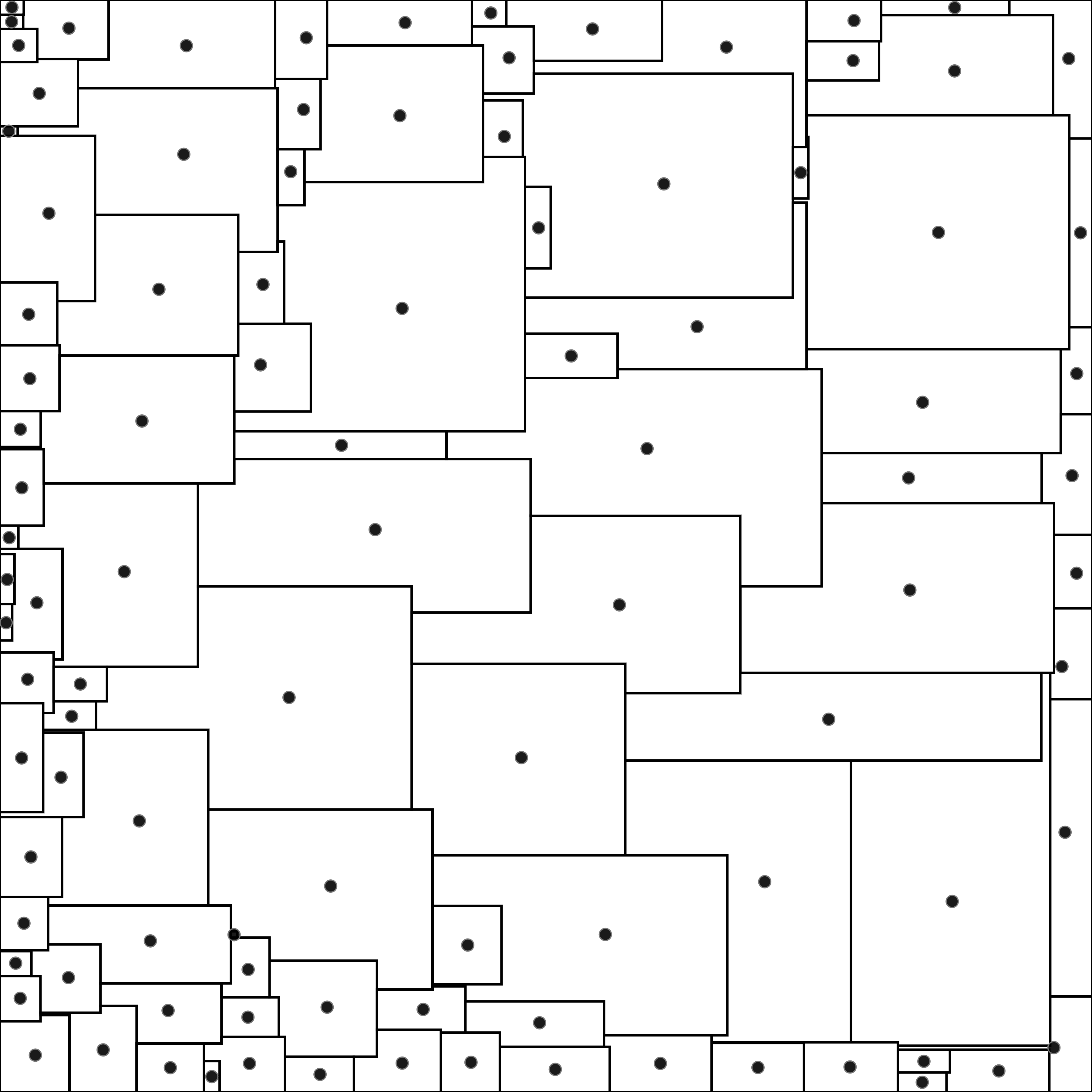} }}%
    \quad
    \subfloat[OVT (Random)]{{\includegraphics[width= 0.4 \columnwidth]{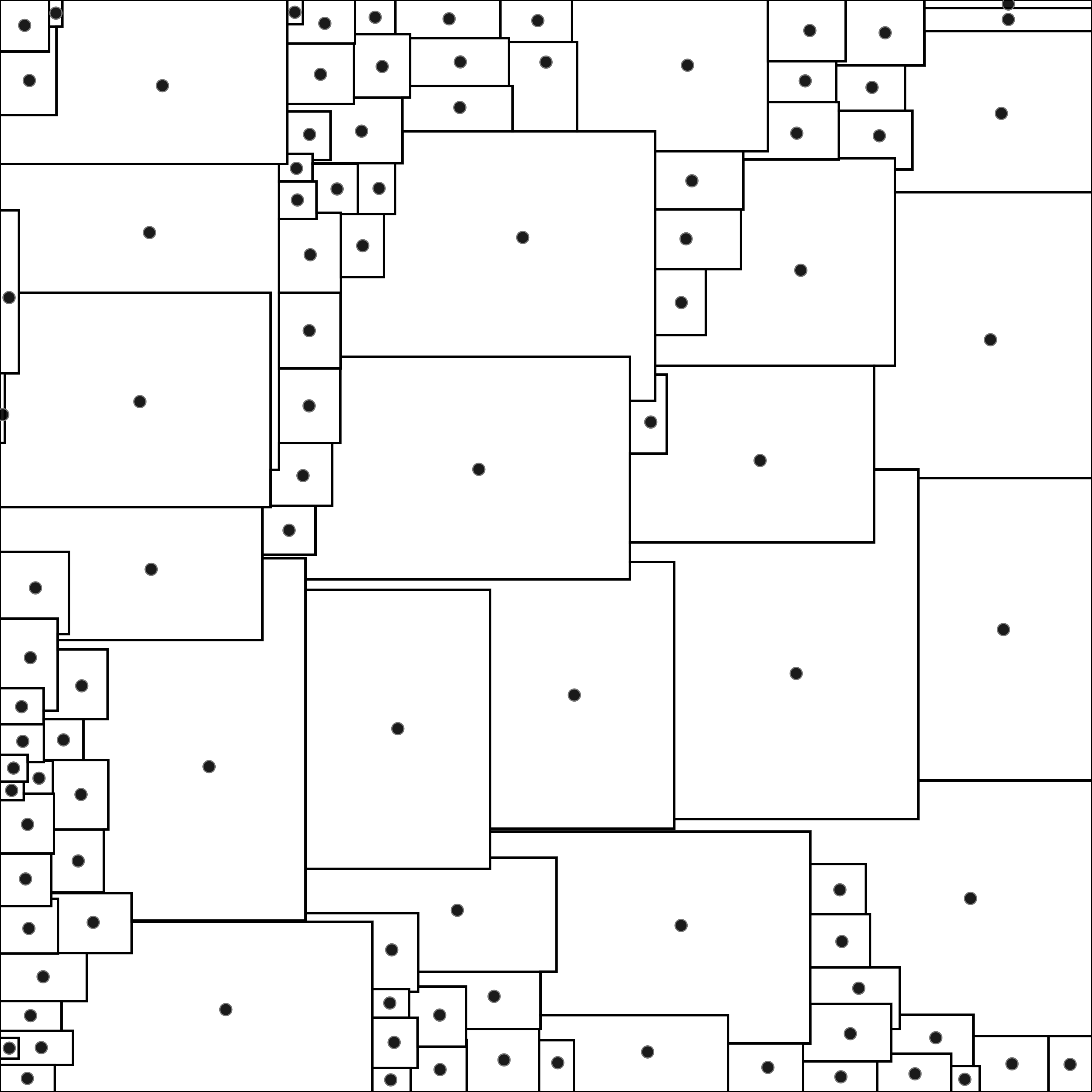} }}%
    \\
    \subfloat[Voronoi Treemap]{{\includegraphics[width= 0.4 \columnwidth]{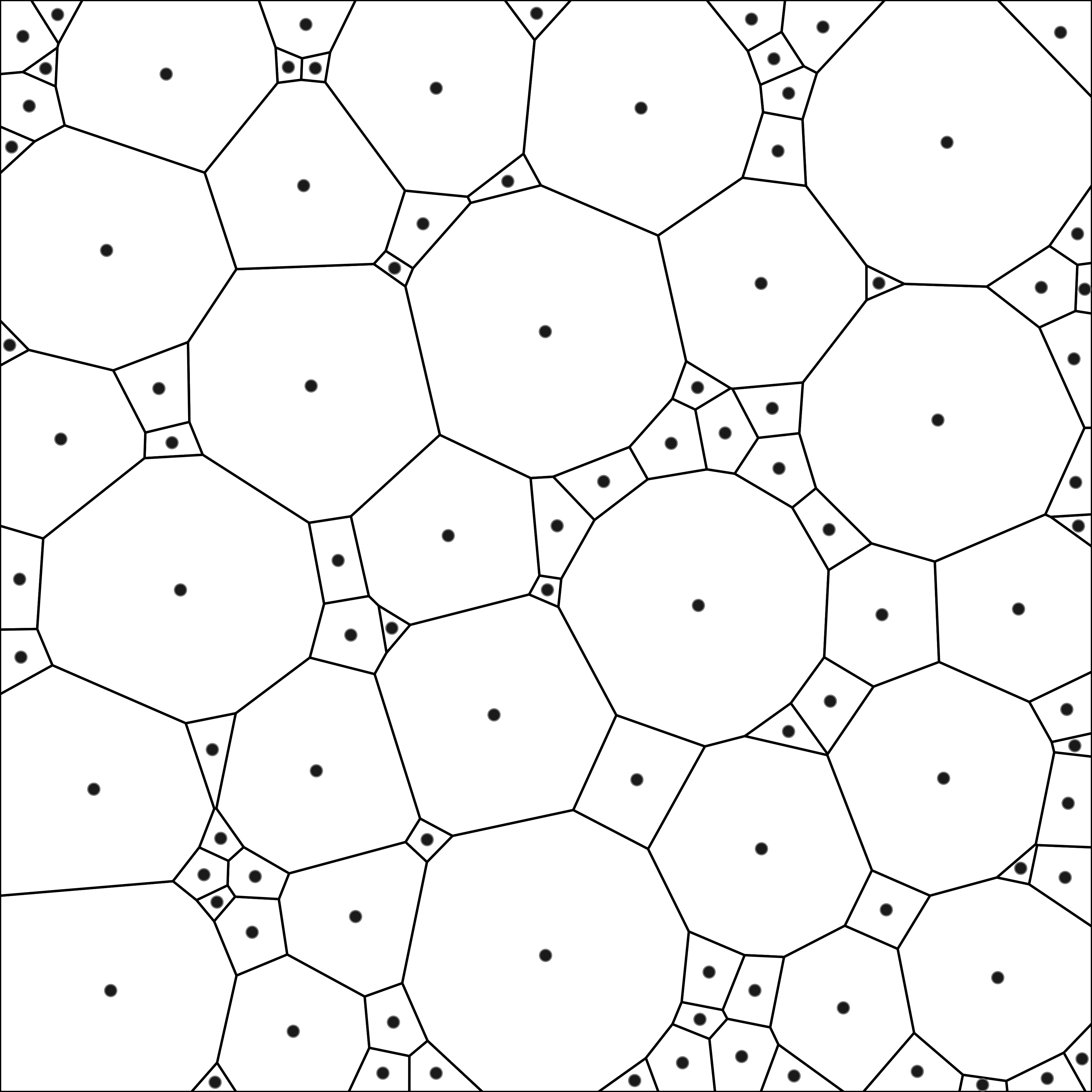} }}%
    %\subfloat[Treemap (Binary)]{{\includegraphics[width= 0.47 \columnwidth]{AspectRatioBinary-1} }}%
    \quad
    \subfloat[Squarified]{{\includegraphics[width= 0.4 \columnwidth]{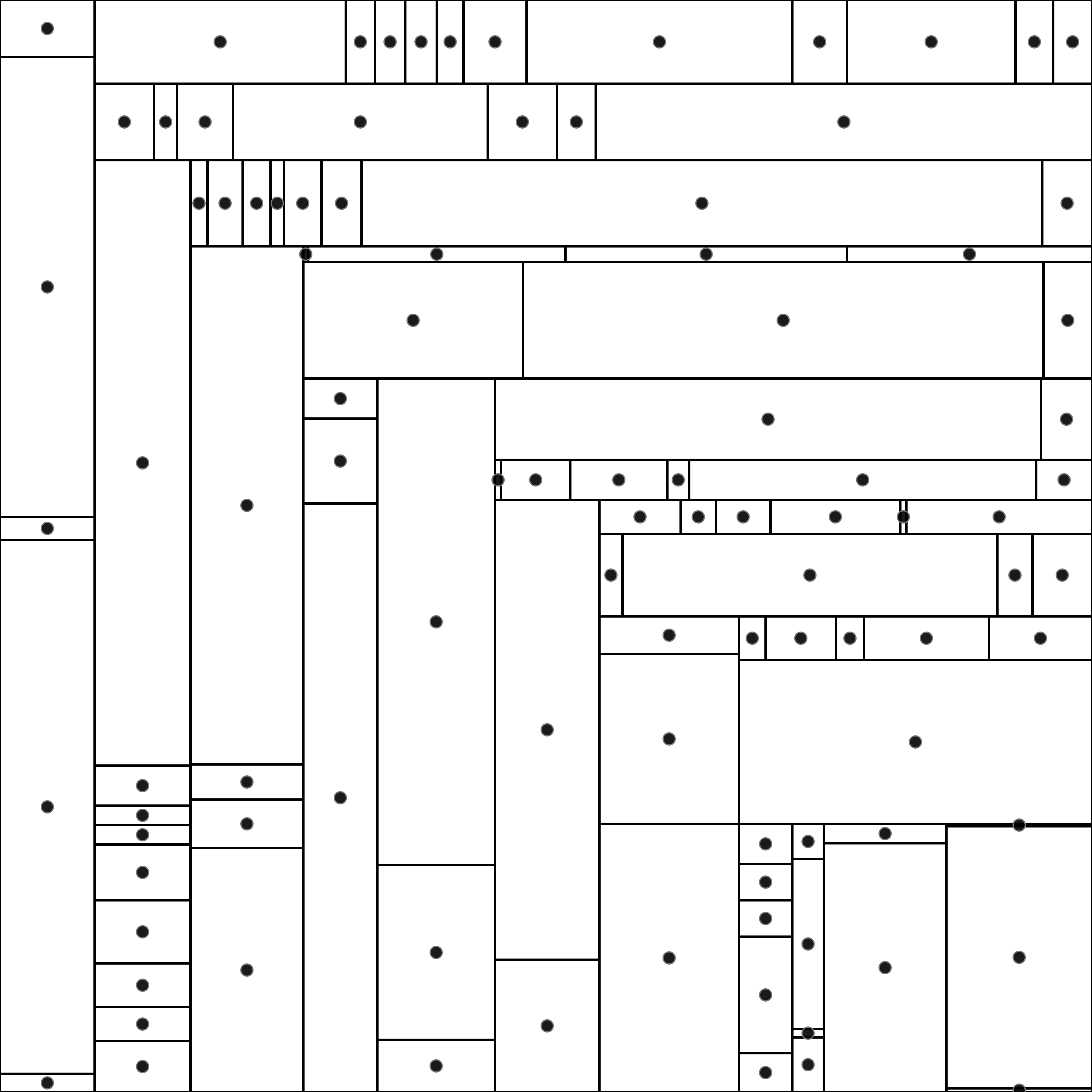} }}%   
    \caption{The final layouts on the random dataset.}%
    \label{fig:exp4-1}%
\end{figure}

%Add Figure Here
\begin{figure}[tb]
    \centering
    \subfloat[OVT]{{\includegraphics[width= 0.4 \columnwidth]{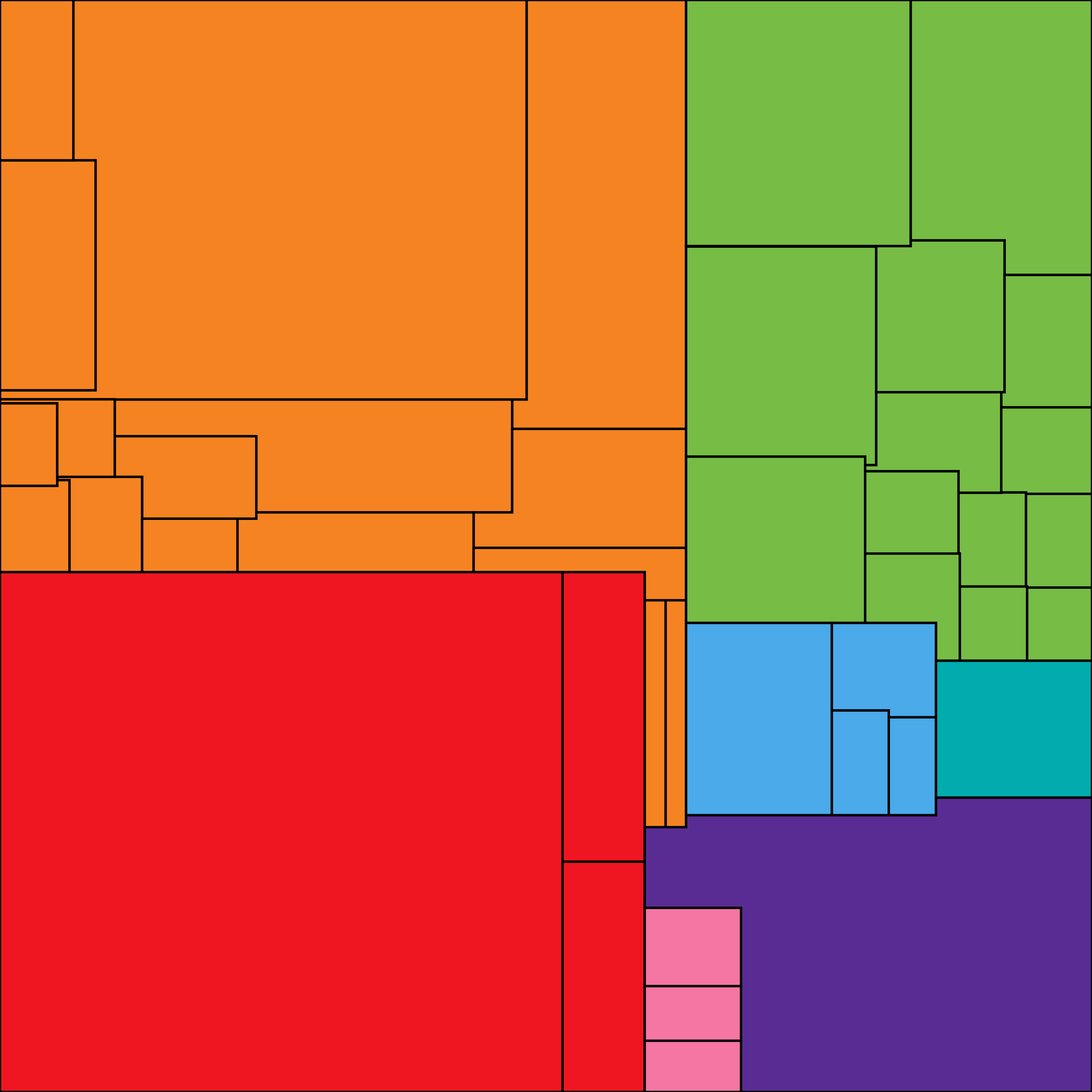} }}%
    \quad
    \subfloat[OVT (Random)]{{\includegraphics[width= 0.4 \columnwidth]{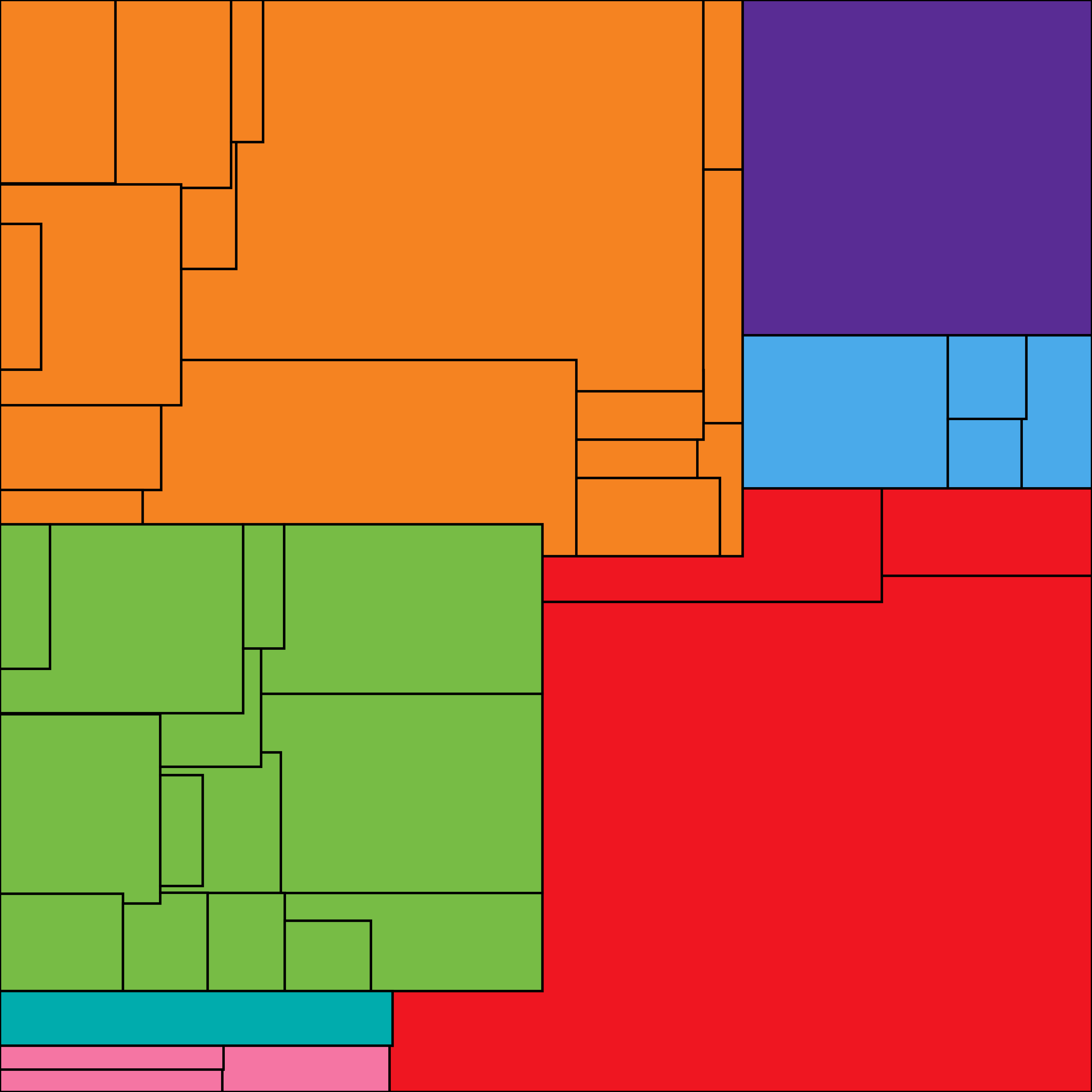} }}%
    %\subfloat[Voronoi Treemap]{{\includegraphics[width= 0.47 \columnwidth]{AspectRatio2-2} }}%
    \\
    \subfloat[Voronoi Treemap]{{\includegraphics[width= 0.4 \columnwidth]{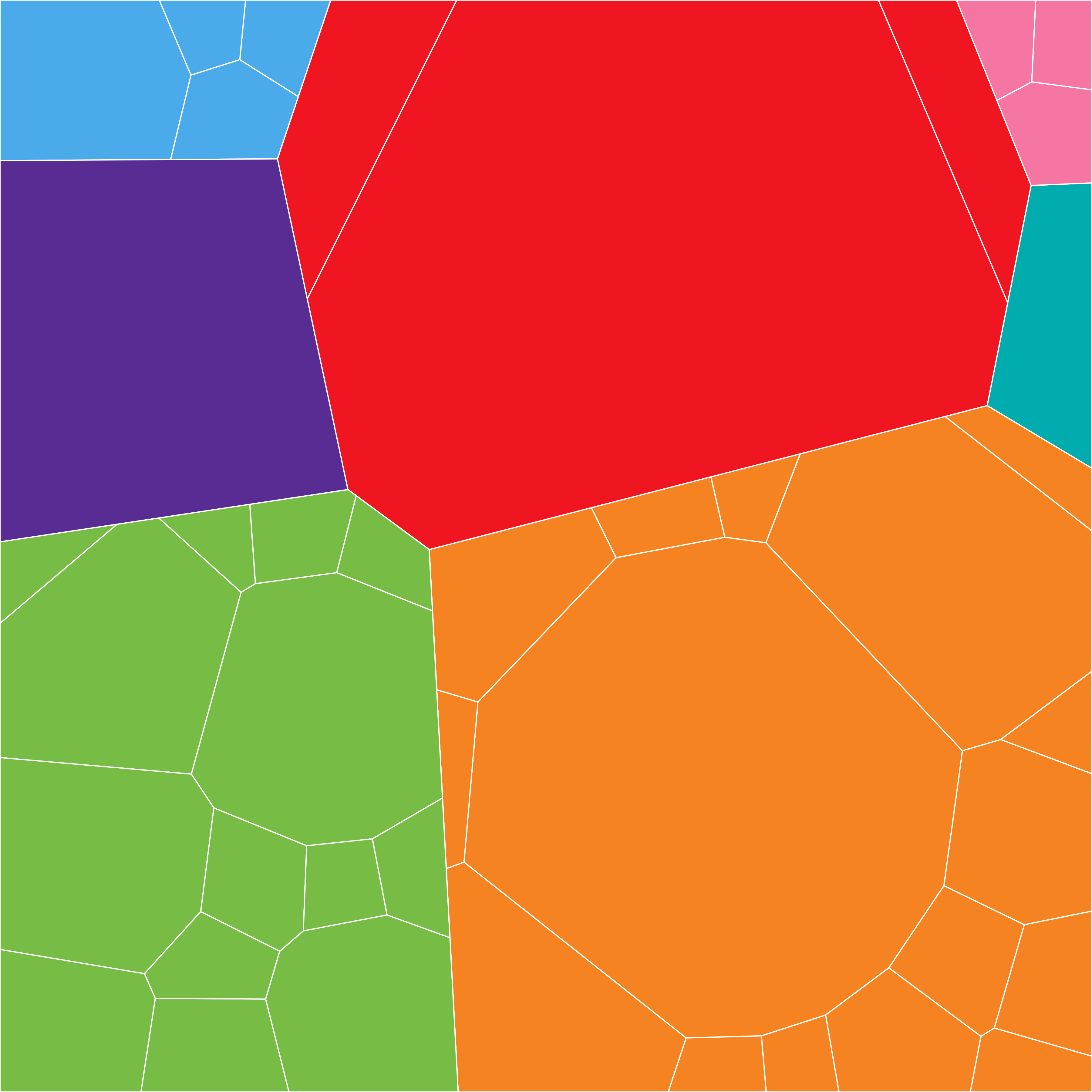} }}%
    %\subfloat[Treemap (Binary)]{{\includegraphics[width= 0.47 \columnwidth]{AspectRatioBinary-2} }}%
    \quad
    \subfloat[Squarified]{{\includegraphics[width= 0.4 \columnwidth]{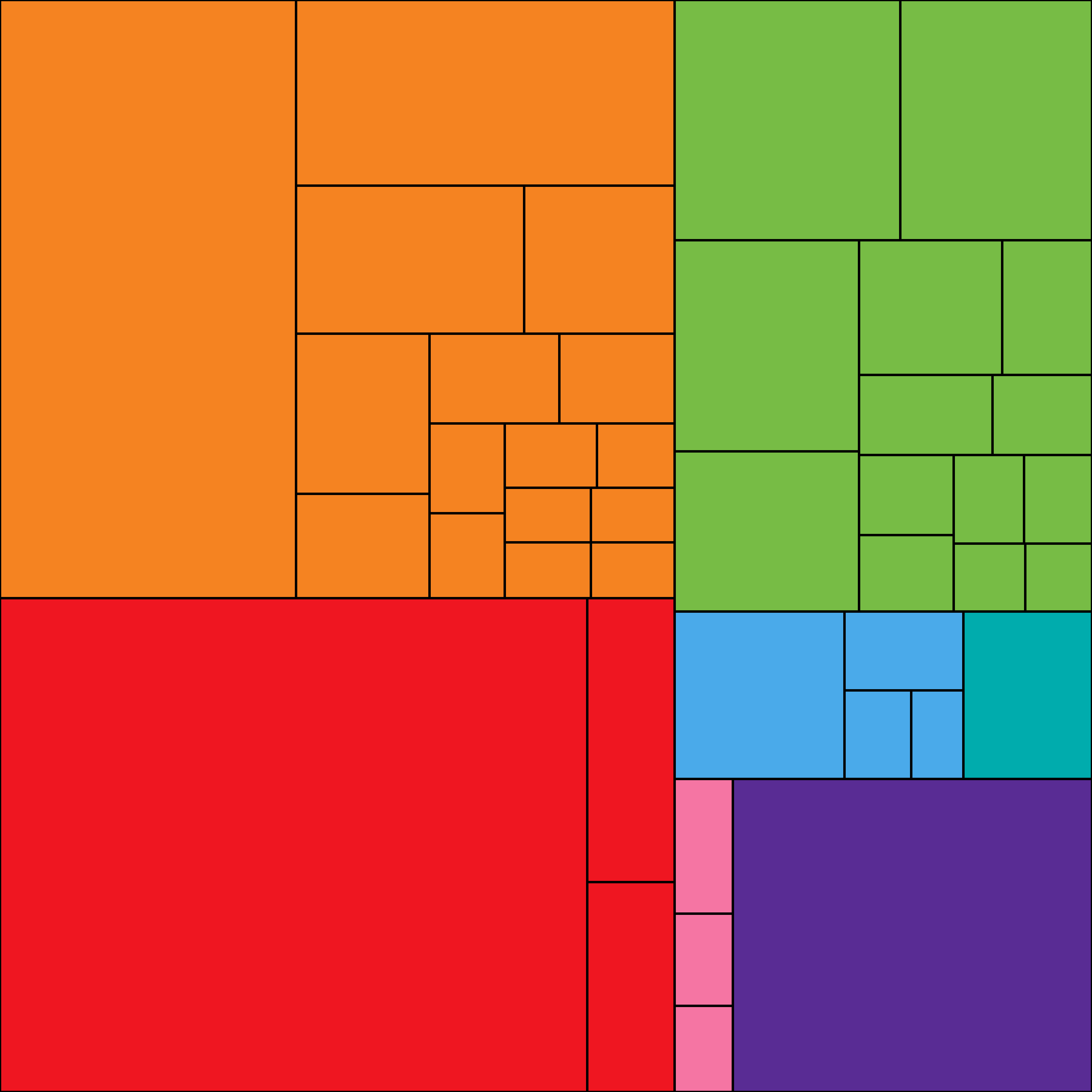} }}%   
    \caption{The final layouts on the globalGDP dataset.}%
    \label{fig:exp4-2}%
\end{figure}

In the following, we evaluate the performance of our proposed algorithm from multiple aspects. First, we check the average number of site pairs and valid neighbors for each site since they have a strong relationship with the algorithm complexity in Sect.~\ref{subsec:exp1}. Second, we compare our algorithm with the Voronoi treemap in terms of running time during a single iteration in Sect.~\ref{subsec:exp2}. Then, the converging rate and the aspect ratio are analyzed in Sect.~\ref{subsec:exp3} and Sect.~\ref{subsec:exp4}, respectively.

\subsection{Site Pairs and Valid Neighbor}
\label{subsec:exp1}

The number of neighbors for each site contributes to the computation complexity of our proposed algorithms. As depicted in Algorithm~\ref{alg:diagram}, all the non-closed sites will be checked to update the status by generating new segmentation lines and updating the skyline. This will lead to an $O(n^2)$ computation complexity. However, in practice, much fewer times are needed for the diagram to update its status. To show this, we generate a series of random datasets with a different number of sites (from $500$ to $50000$). For each dataset, the positions of sites are initialized randomly in an empty square canvas ($900 \times 900$). We run our proposed algorithm on each dataset ten times to have a fair result since our tests are based on random data. During the experiments, two variables are recorded: the total number of site pairs checked and the total number of valid neighbors which change the diagram status. In the end, the average number of site pairs and the average number of valid neighbors per site are illustrated in Fig.~\ref{fig:exp1} as box plots (noted that the x-axis is not linear).

The average number of site pairs per site refers to how many pairs of sites need to be checked for each site. If they are not valid neighbors, then the diagram status will not be changed. Since the check process is much less time expensive than updating the diagram, a small amount of computation time will be taken. It should be noted that the average number of site pairs per site is not linearly increased along the number of sites. We formulate the relationship between the number of sites and the average number of site pairs per site as follows:
\begin{equation}
\#\ of\ site pairs\ per\ site = 0.8563 * 2^{0.7256*log(\#\ of\ sites) } + 0.2494.
\end{equation}
To evaluate the goodness of fit, we calculate the goodness-of-fit statistics including the sum of squares due to error (SSE), R-square, Adjusted R-square, root mean squared error (RMSE)~\cite{Matlab}. In this case, the mean values in the boxplot are used for fitting and we get a good fitting with SSE=0.0664, R-square=1.0, Adjusted R-square=1.0, and RMSE=0.0815.

Although the number of site pairs per site is large, there are only limited valid neighbors which will lead to the change of the diagram status. According to Fig.~\ref{fig:exp1} (bottom), we formulate the relationship by fitting, which can be expressed as:
\begin{equation}
\#\ of\ valid\ neighbors\ per\ site = 0.02875 * log(\#\ of\ sites) + 2.592.
\end{equation}
The goodness-of-fit statistics in this case is SSE=0.0018, R-square=0.8929, Adjusted R-square=0.8832, and RMSE=0.0127 which indicate that the fitting function is suitable. We plot the fit and the prediction in Fig.~\ref{fig:exp1-1}.

Formally, we can conclude that the computation complexity for a single iteration is $O(n\cdot \text{log}(n))$ and the overall complexity is $O(k\cdot n \cdot \text{log}(n))$ in Algorithm~\ref{alg:diagram}. Since the complexity for updating sites status (Algorithm~\ref{alg:adapt}) is linear, then we can show that the proposed orthogonal Voronoi treemap has $O(k\cdot n \cdot \text{log}(n))$ complexity for a single layer hierarchical data.

\subsection{Single Iteration Comparison}
\label{subsec:exp2}
In addition to the computation complexity analysis above, we directly compare the running time of our algorithm during a single iteration to that of the Voronoi treemap. Although several running time are provided in the literature~\cite{Balzer:05, Sud:10, Nocaj:12}, we compare our algorithm with the approach of Nocaj and Brandes~\cite{Nocaj:12} since it is the fastest to date. However, their algorithm is implemented in a different programming language. To have a fire comparison, we utilize the JavaScript implementation of their algorithm instead. We run the JavaScript Voronoi treemap package~\cite{Voronoi} and our algorithm on the same computer (PC, Window 10, Intel Xeon CPU, 3.6 GHz, 16 GB memory). The experiment is also conducted on a series of random datasets with a different number of sites (from 50 to 600) which are associated with random values (approximately 30 \% of sites with random values in $[0,10]$ while the rest sites with random values in $[0,1]$). For each dataset, both algorithms run 1000 iterations (we comment out the converge constraint in Algorithm~\ref{alg:treemap} such that the algorithm will keep running until the maximum iteration number is reached). The average running time (in $ms$) is recorded in Table~\ref{tab:time} and plotted in Fig.~\ref{fig:exp2}. Meanwhile, we list the concrete running time by Nocaj and Brandes~\cite{Nocaj:12} in Java as a reference.  

Our algorithm needs less time per iteration than the Voronoi treemap as shown in Fig.~\ref{fig:exp2} when both are run in the same programming language. The reason for this is that we modified the update sites status process in Sect.~\ref{subsec:update}. In one iteration of the Voronoi treemap, it updates the sites' positions and weights, draws a Voronoi treemap, updates weights and then again draws a Voronoi treemap. In our modification, we merge the update of positions and weights into one step and handle the overweight case in the generation of segmentation line (Algorithm~\ref{alg:line}). Hence in one iteration of our algorithm, we only need to draw the orthogonal Voronoi treemap once rather than twice in the Voronoi treemap.

Compared with the original running time provided by Nocaj and Brandes~\cite{Nocaj:12}, both our algorithm and the Voronoi treemap JavaScript package~\cite{Voronoi} need much more time. Since no hardware-accelerated code used in all cases, we believe that this difference is due to the capabilities of different programming languages. It would be one interesting future work to increase the efficiency by implementing the algorithm in Java in the backend and then displaying the results in JavaScript in the frontend.

\subsection{Converge Rate}
\label{subsec:exp3}
In this section, we test the converging speed of algorithms by considering the changes of the area error ratio along the iteration number increases, since the area of each element in the plot should match its associated value. Starting from initial positions and initial weights, the status of sites is updated iteratively as discussed in Sect.~\ref{subsec:update}. To test the converging speed, we generate a single-layer hierarchical dataset with $200$ random positioned sites which are associated with random values (approximately 30 \% of sites with random values in $[0,10]$ while the rest sites with random values in $[0,1]$). Both the Voronoi treemap and our algorithms are run 5 times. Moreover, we compare our initialization strategy with a random initial status for our algorithms. We calculate and record the area error ratios of both algorithms in each iteration. The results are plotted in Fig.~\ref{fig:exp3}.

As illustrated, our algorithm with designed initial status performs better than that with random initial status and the Voronoi treemap. Compared with random initial status, the proposed initial status guarantees that the algorithm starts from a lower area error. Moreover, the area error ratio can reach a tough constraint (such as 0.01 area error ratio) within 200 iterations. While for our algorithm with random initial status, even after 500 iterations this tough constraint may not be satisfied. Compared with the Voronoi treemap with the random initial position and small positive weight, our algorithm generally converges faster. It is believed that the reason for this result is our reasonable initialization strategy. Since our orthogonal Voronoi treemap is similar to the treemap, using treemap to set initial status significantly contributes to the fast convergence in our algorithm. However, although our algorithm can converge under a tough constraint, it should be noted that our algorithm with both initial statuses has larger fluctuation than the Voronoi treemap. We consider that this is due to the non-consistent distance function used to partition the space (sometimes horizontal, sometimes vertical). While for the Voronoi treemap, the partitioning is straightforward.

\subsection{Aspect Ratio}
\label{subsec:exp4}
We discuss the aspect ratio of our algorithm and compare with the Voronoi treemap and different treemap layouts in this section. Our experiments are conducted on three datasets, including a random dataset (single layer with 100 sites), the global GDP dataset(two layers with 42 leave nodes), and the Flare class hierarchy (four layers with 220 leave nodes). For our algorithm and the Voronoi treemap, we set the maximum iteration number to 500 and the area error threshold for convergence to 0.01. For our algorithm and the treemaps, the aspect ratio of the axis-aligned minimum bounding box is calculated while for the Voronoi treemap, the aspect ratio of the oriented minimum bounding box is calculated. The result is illustrated as boxplots shown in Fig.~\ref{fig:exp4}. Meanwhile, we display the final layouts on the random dataset in Fig.~\ref{fig:exp4-1} and the final layouts on the globalGDP dataset in Fig.~\ref{fig:exp4-2}. The final layout of our algorithm on the Flare data is shown in Fig.~\ref{fig:teaser} (b,c) as well.

When considering the mean aspect ratio (the red line) in Fig.~\ref{fig:exp4}, our algorithm is better than treemap layouts on the random dataset and has similar results on two real datasets, although the Voronoi treemap has the best aspect ratio in three cases. When comparing the different initial status of our algorithm, we can find that with the designed initialization strategy our algorithm has a better aspect ratio (the closer to one the better) and small distribution range. 
%However, our algorithm has large range of aspect ratio distribution compared with other plots. This is mainly due to the random initially positioned sites in our algorithm. For some sites, during the site status update procedure, they are pushed towards the boundaries so that a sub-region with bad aspect ratio will be generated.

\section{Discussion}
\label{sec:discuss}
Our algorithm has a comparable performance with the state-of-the-art Voronoi treemap algorithms in terms of computation complexity, computation time, converge speed, and aspect ratio. Moreover, according to the description of our algorithm, the proposed orthogonal Voronoi treemap is flexible to the changes of data value similar to the Voronoi treemap. Meanwhile, it utilizes nested orthogonal rectangles to present each cell and produces a rectangle-like layout. 

There are also a few drawbacks in our algorithm. Firstly, as mentioned in Sect.~\ref{subsec:exp3}, our algorithm has large fluctuation during the iteration. This shows that the area error in our algorithm is not monotonously decreasing. Fortunately, with our designed initialization strategy, the area error can be decreased to an acceptably small value. The second drawback is that the position of some sites in our algorithm may move in a large range. This leads to a non-stable layout. In the site status update procedure, we have no strategy used to preserve the relative positions of sites. 

It would be an interesting future work for our algorithm to preserve the relative positions of sites during iteration in order to visualize dynamic hierarchical data for which a stable layout is essential~\cite{Sud:10, Hahn:14, Sondag:18}. Another future work is to utilize the treemap to visualize high-dimensional hierarchical dataset. Currently, the parameter associated to each site is usually only one dimension. When it comes to high-dimensional data visualization~\cite{Wang:18}, how to visualize the values or other information while indicating the hierarchical structure would be an interesting direction.

%\cite{}
\section{Conclusion}
\label{sec:conclusion}
In this paper, we described a novel algorithm for the Voronoi treemap with nested orthogonal rectangles. To the best of our knowledge, this is the first time the sweep line strategy is used to generate the Voronoi treemap. We proved that the proposed algorithm has an $O(n \cdot \text{log}(n))$ complexity which is the same as the state-of-the-art Voronoi treemap. Moreover, by modifying the update procedure, it is proved that our algorithm requires less computation time than the Voronoi treemap in the same programming language. Owing to the designed initialization strategy, our algorithm converges faster than the Voronoi treemap and has comparable aspect ratio against the treemap. With a tidy layout by nested orthogonal rectangles and the adjustment capability by site status modification, the proposed orthogonal Voronoi treemap plays the role of the bridge in connecting the treemap and the Voronoi treemap.

%\acknowledgments{The authors wish to thank A, B, and C. This work was supported in part by a grant from XYZ (\# 12345-67890).}

\bibliographystyle{unsrt}

\bibliography{ms}

\end{document}